\documentclass{JINST}
\usepackage{epsfig}
\usepackage{graphicx}

\title{A Multivariate Training Technique with Event Reweighting}

\author{Hai-Jun Yang$^*$, Tiesheng Dai, Alan Wilson, Zhengguo Zhao, Bing Zhou \\
Department of Physics, University of Michigan, Ann Arbor, MI 48109-1120, USA \\
$^*$E-mail: \email{yhj@umich.edu}}

\abstract{
An event reweighting technique incorporated in multivariate
training algorithms has been developed and tested with 
Artificial Neural Networks (ANN) and Boosted Decision Trees (BDT). 
The performance of the ANNs and BDTs resulting from this event reweighting training is compared to the 
performance from conventional equal event weighting training.
The comparison is performed in the context
of physics analysis in the ATLAS experiment at the 
Large Hadron Collider (LHC), which will explore the fundamental nature of 
matter and the basic forces that shape our universe.
We demonstrate that the event reweighting technique provides an
unbiased method of multivariate training for event pattern recognition. 
}


\begin{document}

\section{Introduction}

Artificial Neural Networks (ANN) and Boosted Decision Trees (BDT)
\cite{tmva,bdt1,bdt2,bdt3,adaboost,eboost} 
are two important data analysis tools that have wide application in High Energy 
Physics experiments for particle identification and for
event pattern recognition~\cite{a1,a2,a3,a4}. Both methods 'train' the 'networks' or the 'trees'
based on a set of 'signal' and 'background' features (physical quantities)
to obtain a powerful discriminant variable that distinguishes signal from background. 
This process is called 'event pattern recognition' in physics data analysis.
In the conventional ANN and BDT algorithms for high energy
physics analysis, the training events (including signal and background) 
are initialized with equal weights.
The equal event weight training technique works fine if the 
Monte Carlo (MC) samples from different
physics processes used for training are generated based on 
their production rates. In physics studies, we need to have very
large MC 'background' event samples to determine the rates of 
the misclassified events from different processes that contaminate
the signal (normally a few times more MC data than real data is needed for a 
certain integrated luminosity). 
However, for hadron colliders such as the Large Hadron Collider at CERN
(LHC, http://cern.ch/) and Tevatron at Fermilab (http://www.fnal.gov/),
it is unrealistic and inefficient to generate MC data 
for all the physics processes with full detector simulations based on 
their production rates. This is simply 
because of limited CPU time and data storage capacity. To simulate and reconstruct an event for the ATLAS
experiment, typically, it would take about 10 minutes of CPU time
and about 2.5 MB of storage space
per event. The simulation time is many orders of magnitude longer
than the event created from the beam collisions. 

Combining statistically limited MC events from different physics processes 
raises a natural question on the multivariate training process in event weighting.
Suppose that 100K MC events are generated for each of background A and background B.  
Suppose, in addition, we expect 80\% of the background to be from A and 
20\% from B.  What is proposed in the event reweighting technique is 
that the events are reweighted for training, so that 80\% of the 
total weight is from A and 20\% is from B, i.e. each event from A has 4 times the 
weight as an event from B. We implemented this idea in ANN and BDT
training programs and tested this technique in the context of the ATLAS experiment with
fully simulated MC datasets. 

ATLAS (http://atlas.web.cern.ch/) is one of two general purpose detectors
at the LHC (a 27 km circumference particle accelerator that will collide
protons head-on with a center of mass energy of 14 TeV) 
being built at the European Center for Nuclear Research (CERN) in Geneva, 
Switzerland. The ATLAS experiment is designed to 
search for signals that would respond to the electroweak symmetry breaking.
Some theoretic models predicted signals such as 
standard model Higgs bosons, supersymmetric particles, and new bosons from
extra dimensions. Discovering any one of those signals at the LHC would
be a great breakthrough in our understanding of particle physics. The LHC
will begin operation in 2008. A major part of preparing for
LHC physics analysis is to develop and test advanced data analysis tools.

In this paper, we use ATLAS MC samples for $WZ \rightarrow \ell\nu \ell\ell$ analysis 
to demonstrate  that using event reweighting technique will provide an unbiased training
in ANN and BDT multivariate analysis.  
Our 'signal' is from the $WZ$ triple-lepton decay channels ($eee\nu$,
$ee\mu\nu$, $\mu\mu e\nu$ and $\mu\mu\mu \nu$). Major backgrounds
come from standard model processes such as $t \bar t$, $Z+jets$, $ZZ$ and $Drell-Yan$.
Those backgrounds have production rates 3-4 orders of magnitude larger than 
that of the signal process. 
Our goals are to maximize signal efficiency, to minimize background
efficiency and to understand the uncertainties with limited training and test samples.

For comparing the performance of the ANN and the BDT with or without event reweighting 
training, we used the same testing sample (statistical independent of training
sample). The main purpose of this paper is to compare the training 
performance with and without event reweighting. 
Performance comparisons between ANN and BDT can be found in
the contexts of MiniBooNE neutrino oscillation analysis\cite{bdt1,bdt3}, 
D0 single top discovery\cite{a3} and B-tagging\cite{a4}.

In section 2 we provide the MC signal and background information, including
the physics processes, the production cross-sections at the LHC, the total simulated
MC event size and the training sample size after pre-selection.
We also give brief descriptions of physics variables for both the ANN and the BDT analysis.
The event reweighting training techniques for BDT and ANN
are presented in section 3. Performance comparisons for different weighting
methods are summarized in section 4.
Section 5 presents uncertainty study results and section 6
gives our conclusions.

\section{MC Samples and Training Variables}

Monte Carlo samples used in this study are from the ATLAS Computing
System Commissioning (CSC)~\cite{CSC} with full detector simulation and reconstruction.
For this study we used a few loose cuts to pre-select
events with the approximate experimental signature of the signal, then the pre-selected events are analyzed using
the ANN and BDT multivariate programs to further separate signal from background events.
In Table~\ref{tab:mc} we list the MC and pre-selection information for the signal ($WZ$) 
and background MC events used in this study.
This information includes the total production cross-sections ($\sigma_{MC}$)~\cite{mc},
the triple-lepton decay branching ratio ($Br$), the total number
of simulated MC events ($N_{MC}$), the number of pre-selected
events ($N_{precut}$), and the number of expected events ($N_{exp}$) normalized
to 1~fb$^{-1}$ integrated luminosity ($\int L dt $)
after the pre-selection.
The initial event weight for each process is listed in Table~\ref{tab:mc} as well.  
The cross-section correction factor, $K$, in the Table~\ref{tab:mc}
is defined as the ratio of the next-to-leading order (NLO) cross-section to the
cross-section obtained from the MC generators.  Thus, if we used
a NLO MC generator, the K value is 1. Otherwise, if a LO MC generator is used,
the K value is $\sigma(NLO) / \sigma(LO)$ ($\sigma $ denotes the cross-section).
The expected number of events for 1~fb$^{-1}$, $N_{exp}$, and the event's {\it weight}
listed in Table~\ref{tab:mc} are calculated based on
$$
N_{exp} = {\sigma_{MC} \times K \times Br \times (\int Ldt) \times N_{precut} \over N_{MC}}
$$ 
and
$$
Weight = {\sigma_{MC} \times K \times Br \times (\int Ldt) \over N_{MC}}.
$$
The integrated luminosity ($\int Ldt$) is a constant (1~fb$^{-1}$) 
for all the MC process, so the event weight of a given MC process depends on 
its production cross-section, decay branching ratio 
and the total number of MC events generated ($N_{MC}$).
For a given MC process, a larger event weight means lower statistics
for analysis. In general, higher MC statistics are desired to reduce
statistical uncertainty for data analysis. 
Evidently, the event weights vary dramatically among the various MC processes as
shown in Table~\ref{tab:mc}.

The pre-selection applied loose cuts to all datasets by requiring
two leptons with invariant mass consistent with the $Z$ mass and 
an additional lepton with missing transverse energy forming a transverse 
mass consistent with the W boson mass. The pre-selection also requires that at least one lepton
have transverse momentum greater than 20~GeV to satisfy the trigger
requirements of the experiment.  

The physics variables input into the ANN and BDT trainings are selected
based on our our experience with cut-based analysis (optimized to separate signal
from background), and the variable 'Gini' index determined from the decision trees
~\cite{bdt1}. This index indicates the separation power between signal and background of a variable.
We give brief descriptions of these variables, separated into
four categories.
\begin{itemize}
\item {\bf Energy and Momentum}\\ 
   The characteristics of energy and momentum are different from the WZ events and the 
background events. For example, $t\bar t$ events will have larger hadronic jet energies 
compared to the $WZ$ events, the $Z+X$ background will have lower missing transverse energy 
and so on. The energy and momentum variables we used are
   \begin{itemize}
    \item $P_T^{\ell}$ - lepton transverse momentum, three variables (two leptons decay from Z, one lepton 
decays from W),
    \item $MET$ - missing transverse energy,
    \item $MET$ significance which is defined as $MET/\sqrt{\sum_{i} E_T(i)}$, (i is the index counting leptons and jets),
    \item $E_T^h$ - vector sum of transverse momentum from leptons and $MET$,
    \item $H_t$ - scalar sum of transverse momentum from jets, leptons and $MET$,
    \item $\sum E_T^{jet}$ - sum of transverse energy from each jet,
    \item $P_T(WZ)$ - transverse momentum of $WZ$ bosons,
    \item $E_T^{recoil}$ - total recoil transverse energy.
   \end{itemize}
\item {\bf Lepton isolations }\\
   The leptons from the $W$ and $Z$ decays are isolated, but leptons from the QCD jets are not isolated.
   Typically, the QCD jets have multiple tracks and larger energy deposition around the leptons. The
   isolation variables we used are
   \begin{itemize}
    \item $N_{trk}^{iso}$ - number of charged tracks in $\Delta R < 0.4$ cone around a lepton
     ($\Delta R = \sqrt{(\Delta\phi)^2 + (\Delta\eta)^2} $), three variables (two leptons decay from Z, one 
lepton decays from W),
    \item $\sum P_T^{iso}$ - sum of track $P_T$ in a $\Delta R < 0.4$ cone around a lepton,
    \item $\sum E_T^{iso}$ - sum of jet transverse energy in $\Delta R < 0.4$ cone around a lepton,
    \item $f_\ell^{iso}$ - fraction of energy = [E($\Delta R < 0.4$)-E($\Delta R < 0.2$ )] / $E_T^{\ell}$.
   \end{itemize}
\item {\bf Event topologies }\\
   The following variables are selected to suppress top and QCD jet events and to separate
   fake lepton events from the $WZ$ signal:
   \begin{itemize}
     \item $\Delta R (\ell,\ell')$ - separation between two leptons, 
2 variables (one lepton decays from W, the other lepton decays from Z),
     \item $\Delta A $ - vertex difference between leptons in transverse plane (impact parameter), 
2 variables (one lepton decays from W, the other lepton decays from Z).
   \end{itemize}
\item {\bf Mass information }\\
   The mass information is used to reduce QCD and $t\bar t$ background events with leptons (or fake leptons) 
   that do not decay from $Z$ and $W$. Two variables are used in our analysis:
   \begin{itemize}
     \item $M_{\ell\ell}$ - invariant mass of two leptons from $Z$ decays,
     \item $M_T(\ell, MET)$ - transverse mass of a lepton and $MET$ (neutrino) from $W$ decays.
   \end{itemize}
\end{itemize}

\begin{figure}
\epsfig{figure=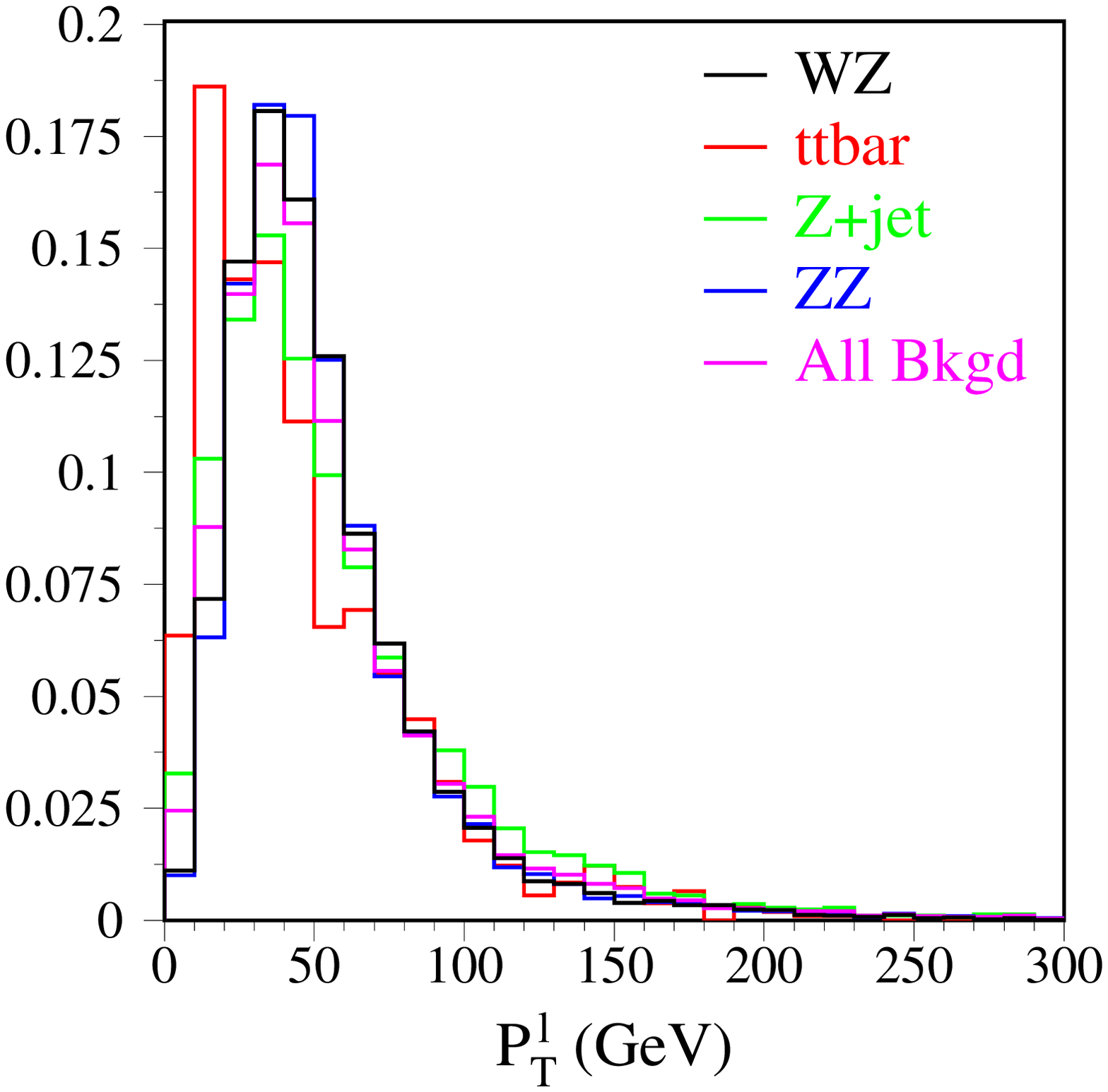,width=8.5cm}
\epsfig{figure=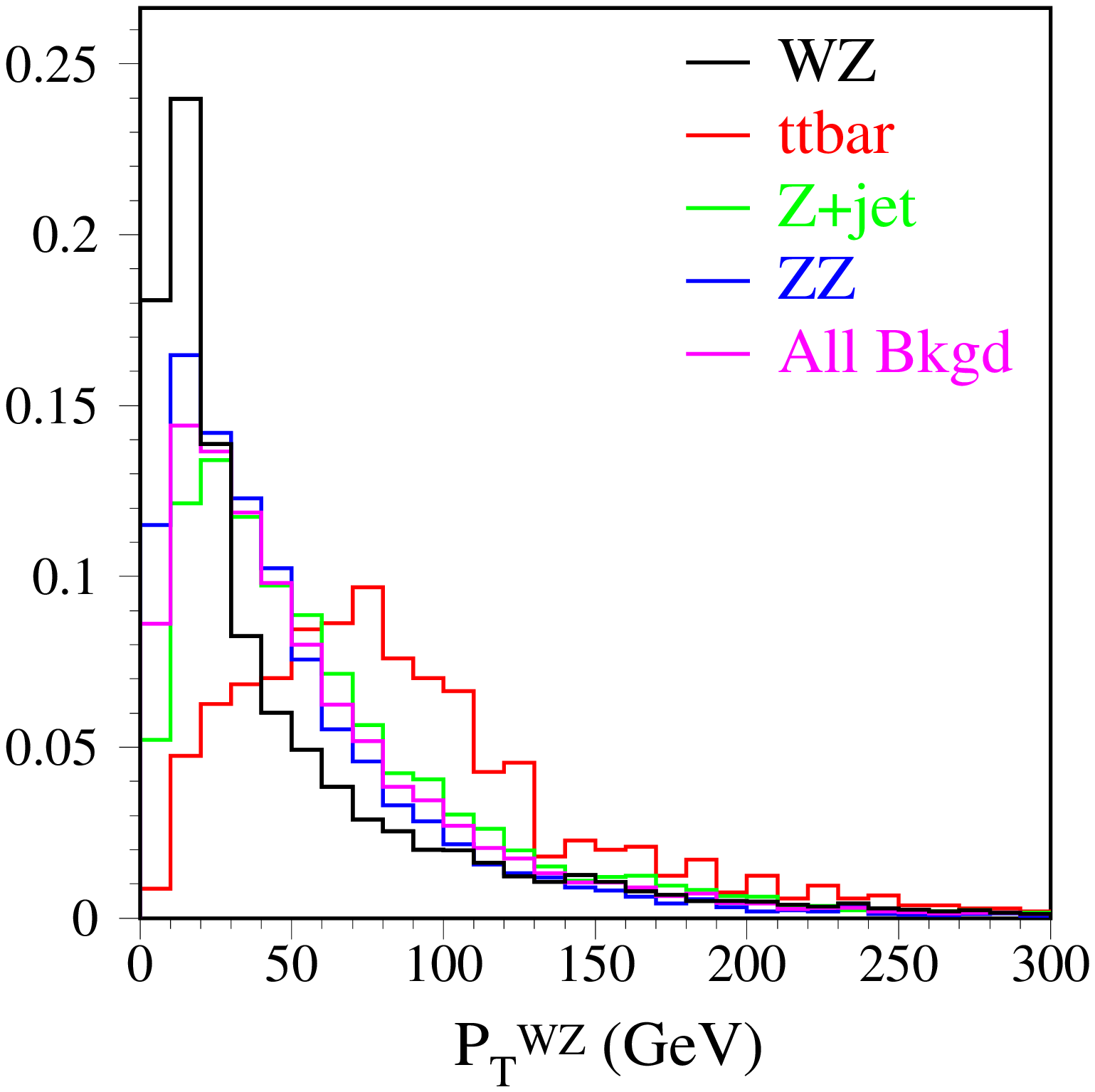,width=8.5cm}
\epsfig{figure=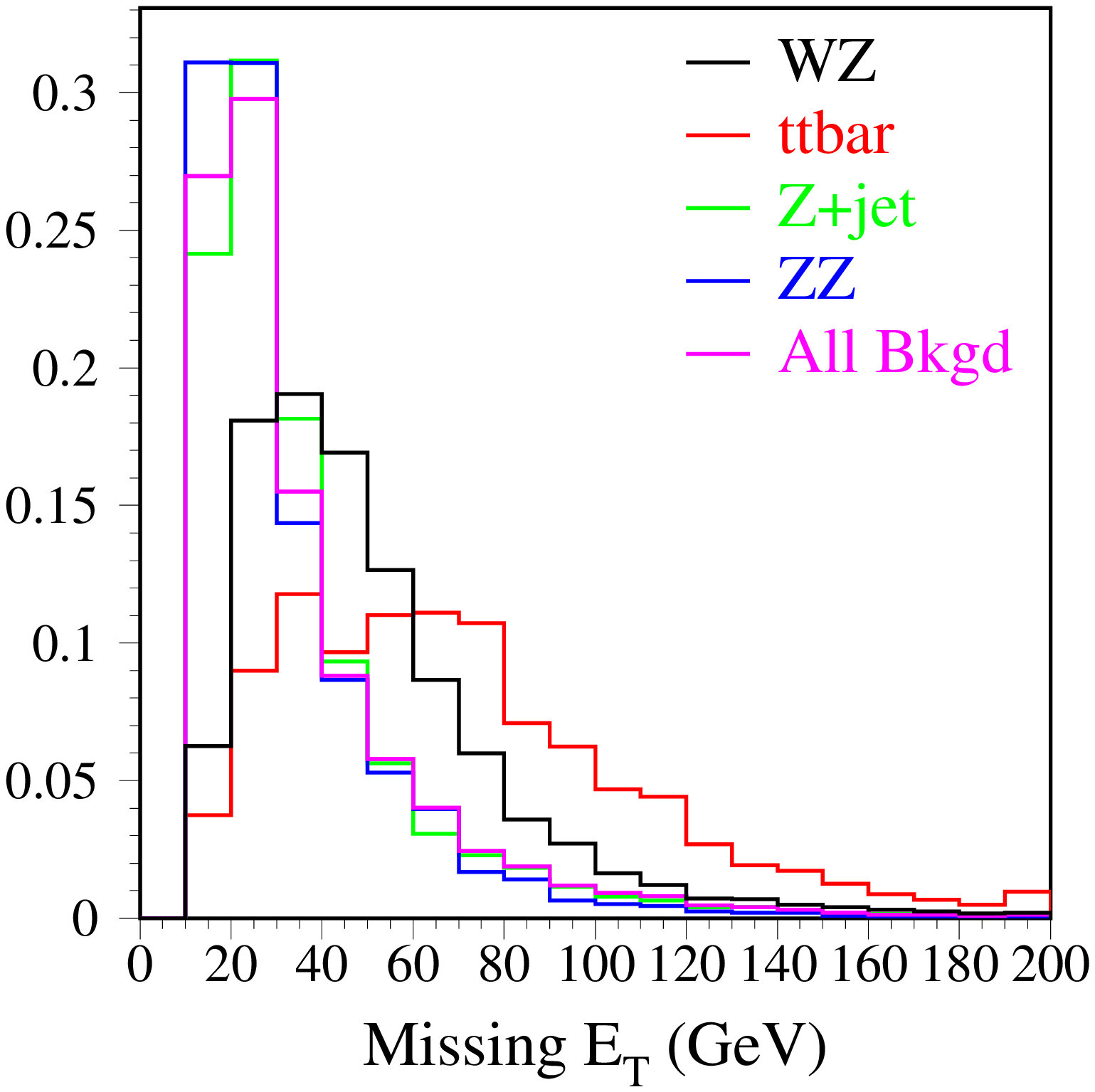,width=8.5cm}
\epsfig{figure=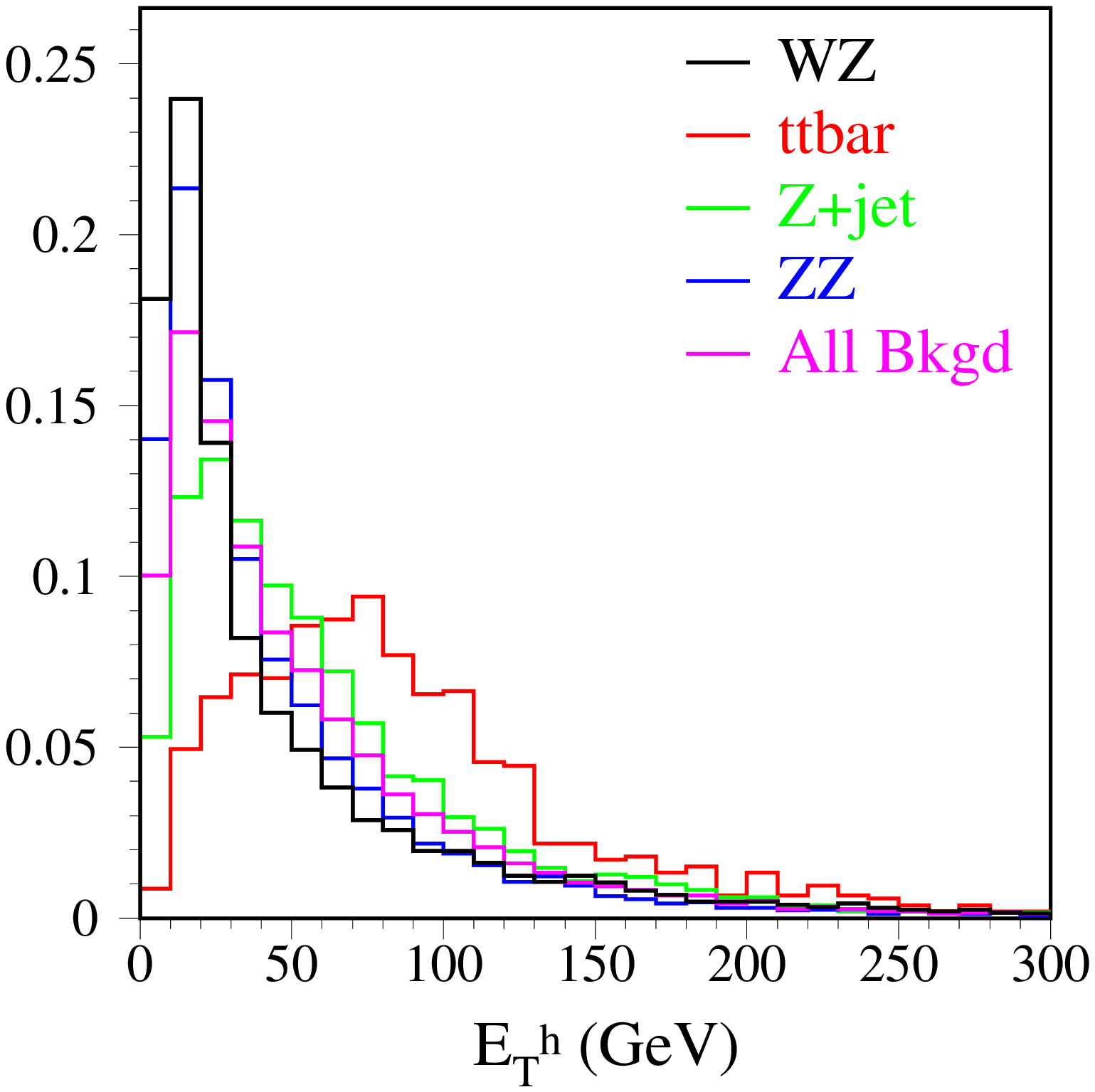,width=8.5cm}
\caption{Distributions of 
the transverse momentum of leptons (top left),
the transverse momentum of the WZ system (top right), 
the missing transverse energy of the event (bottom left) and
the vector sum of transverse momenta from leptons and $MET$ (bottom right).
Among the histograms, black indicates ZW signal events, 
red indicates $t\bar t$, 
green indicates Z plus jets, 
blue indicates $ZZ \rightarrow \ell\ell\ell\ell$ and 
pink indicates a combination of all backgrounds. 
All histograms are normalized to the same area for comparison.}
\label{fig:var1}
\end{figure}

\begin{figure}
\epsfig{figure=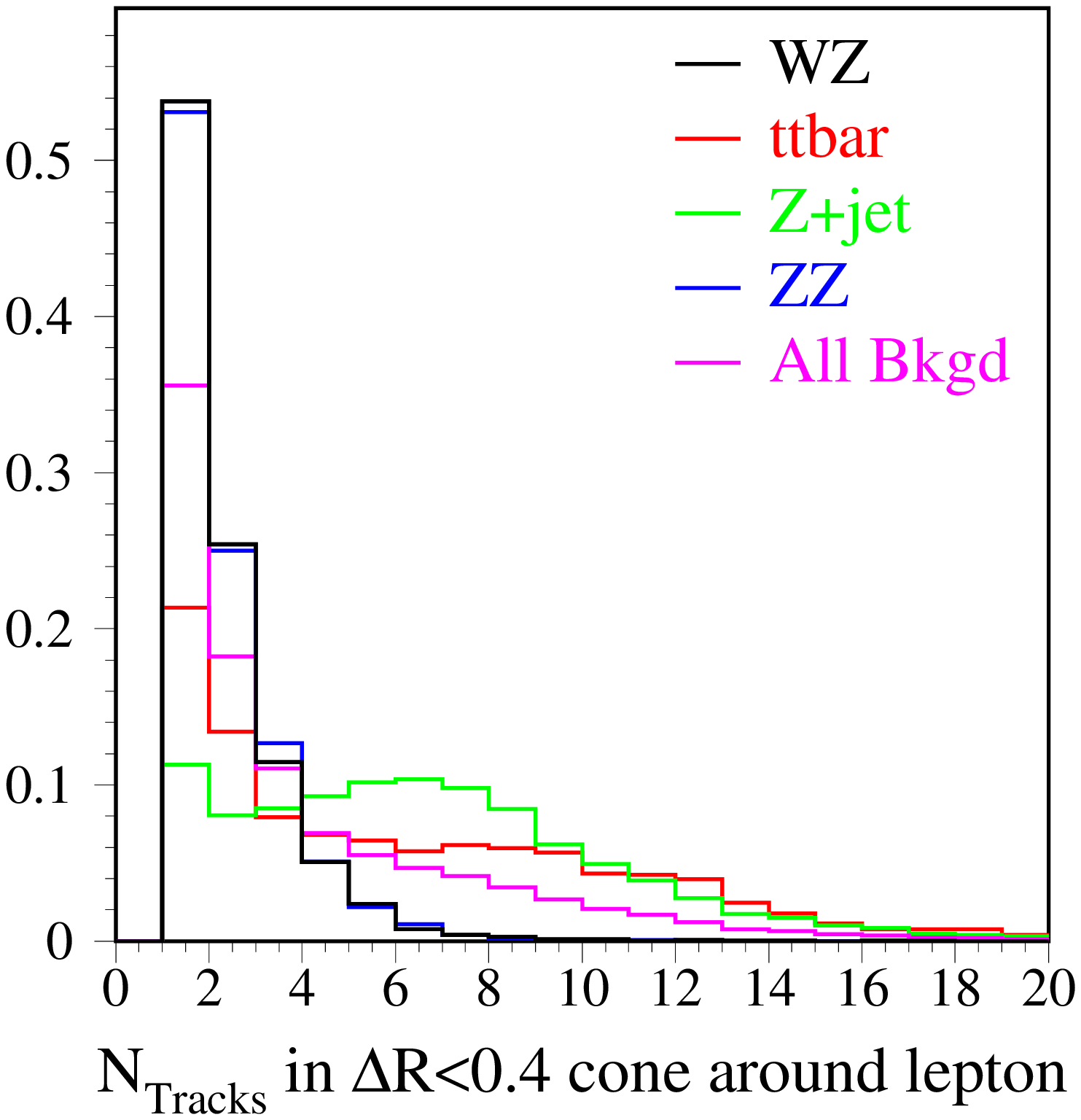,width=8.5cm}
\epsfig{figure=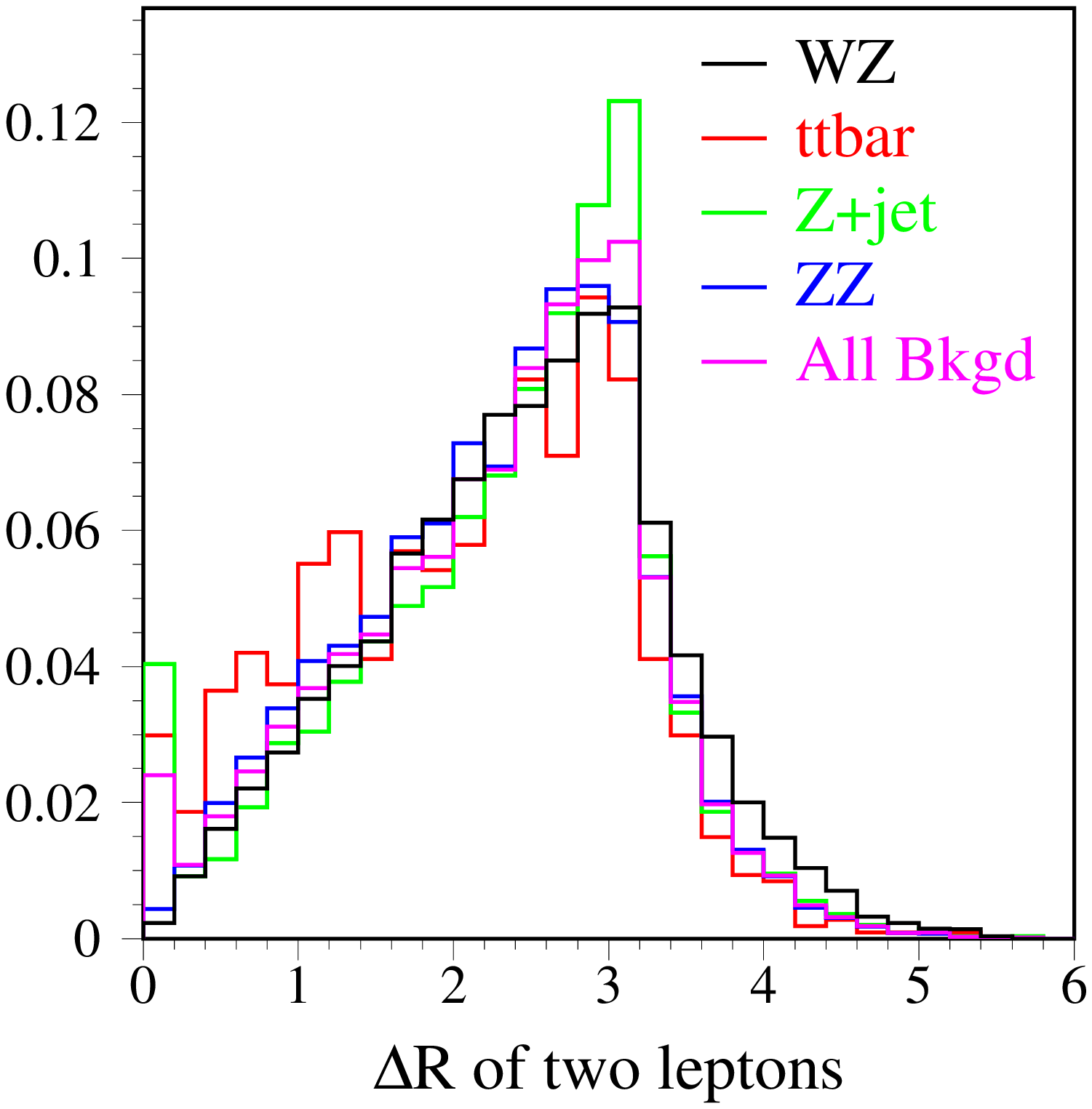,width=8.5cm}
\epsfig{figure=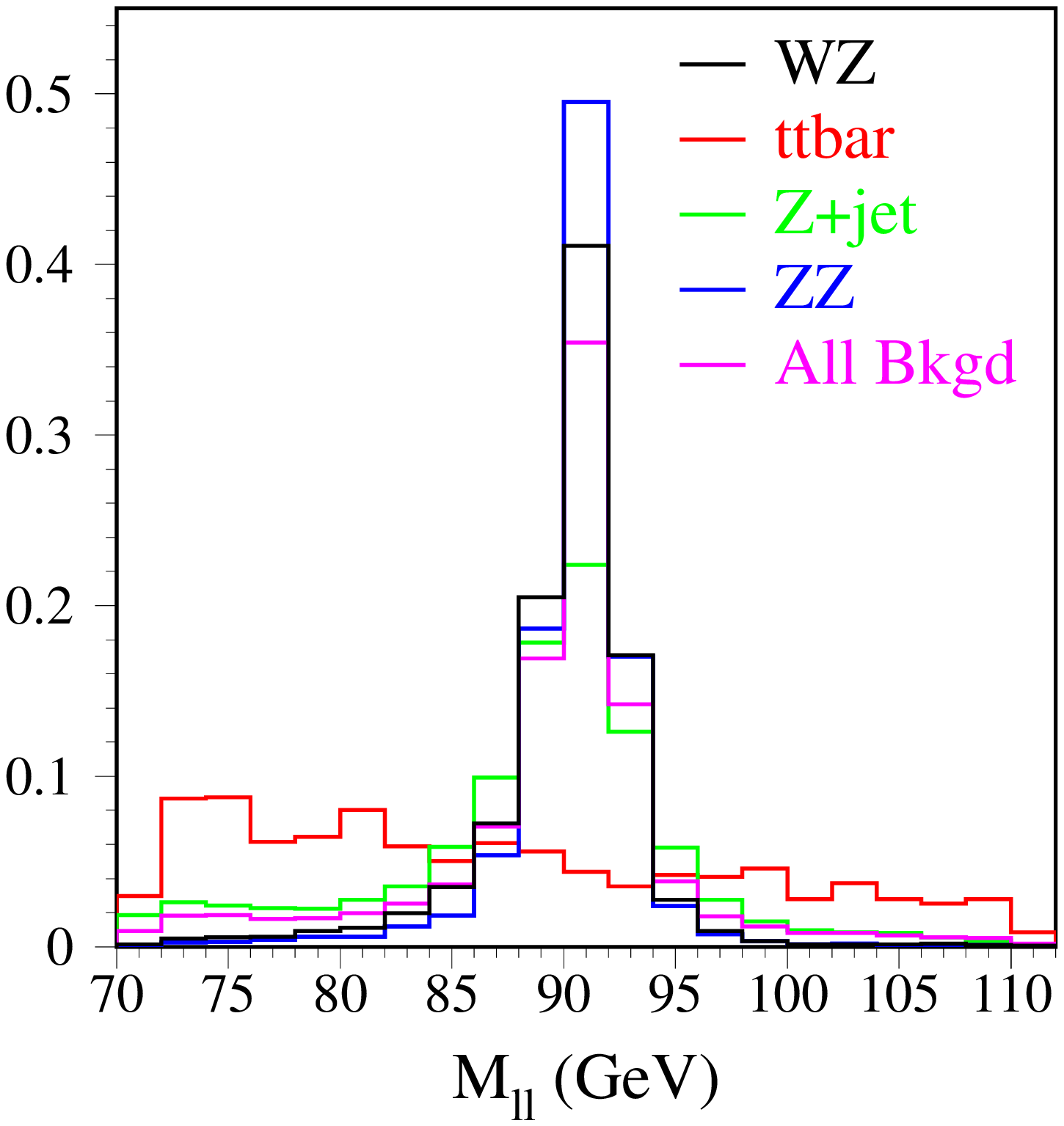,width=8.5cm}
\epsfig{figure=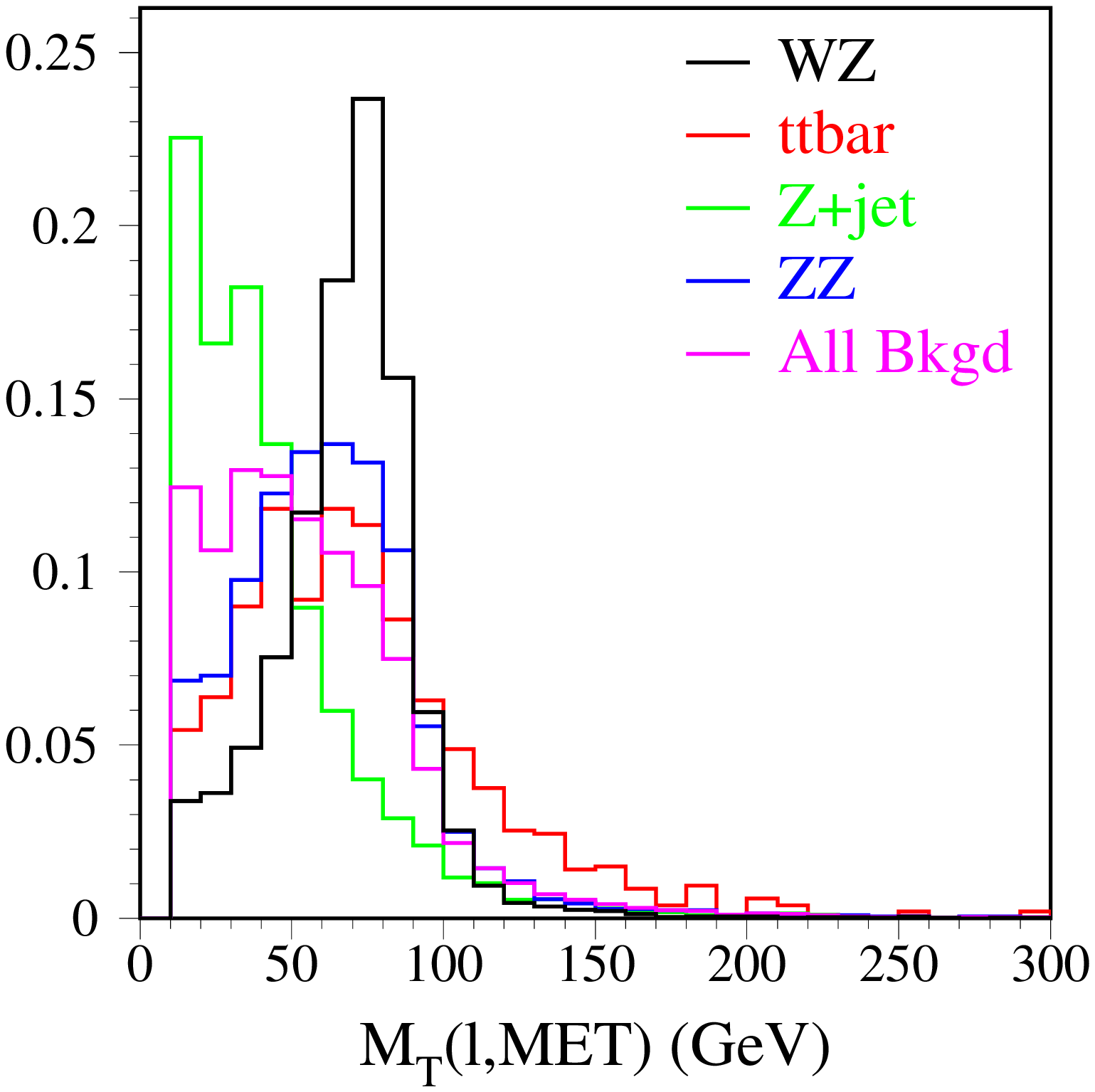,width=8.5cm}
\caption{Distributions of the number of charged tracks around a lepton in a cone of $\Delta R=0.4$ (top left), 
the two lepton separation in $\Delta R$ (top right),
the invariant mass of two leptons (bottom left) and the transverse mass of leptons combined with $MET$ (bottom right).
Among the histograms, black indicates ZW signal events,
red indicates $t\bar t$, 
green indicates Z plus jets, 
blue indicates $ZZ \rightarrow \ell\ell\ell\ell$ and 
pink indicates a combination of all backgrounds. 
All histograms are normalized to the same area for comparison.}
\label{fig:var2}
\end{figure}

For variable distribution shape comparisons,
we show some energy and momentum variable distributions for signal and background 
in Figure~\ref{fig:var1} and some variable distributions related to lepton isolation, event topology
and mass in Figure~\ref{fig:var2}.
All the events in the plots are passed pre-selection cuts and each signal (black histograms) 
and background distribution is normalized to the same area.
  
\begin{figure}
\epsfig{figure=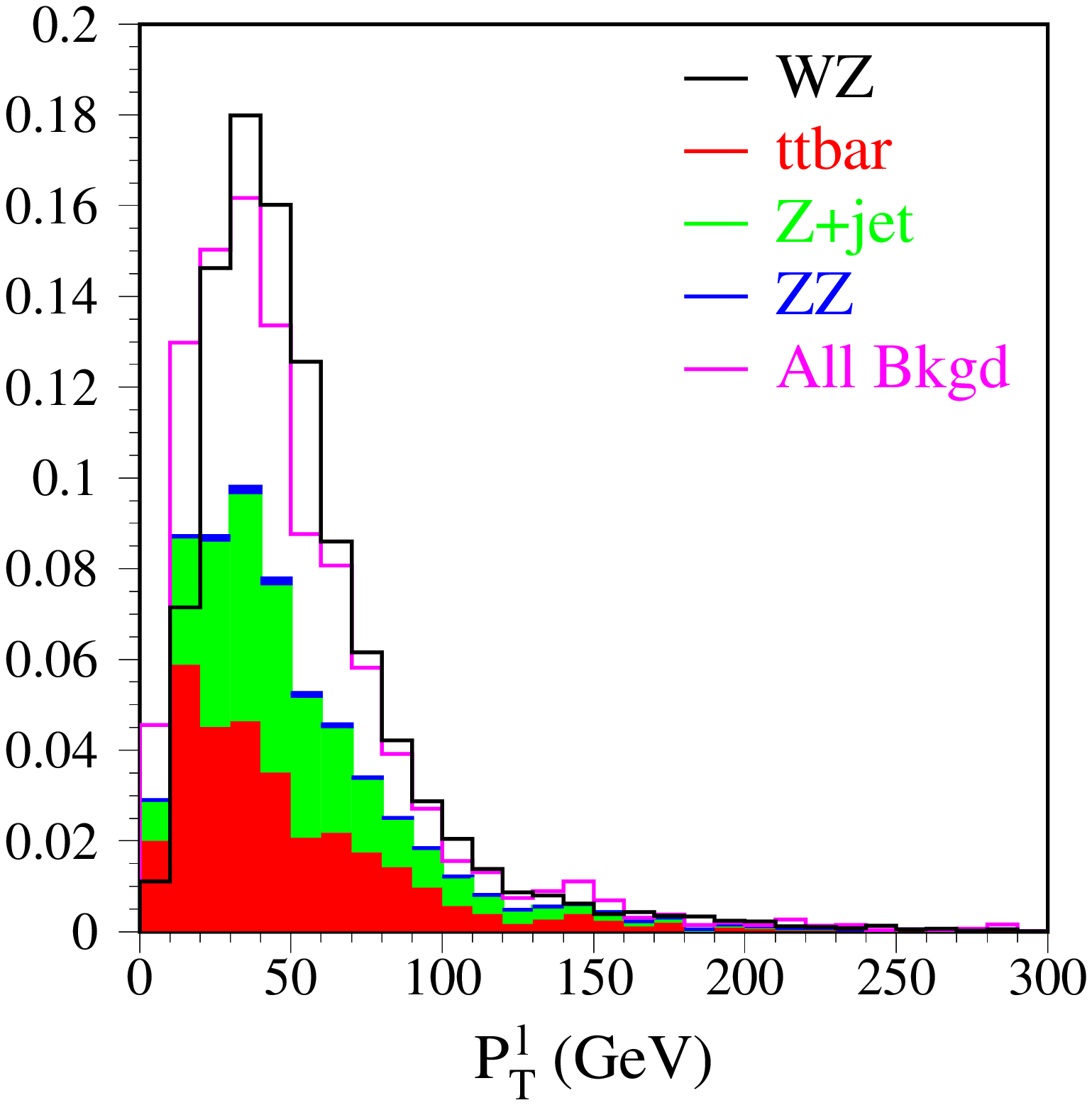,width=8.5cm}
\epsfig{figure=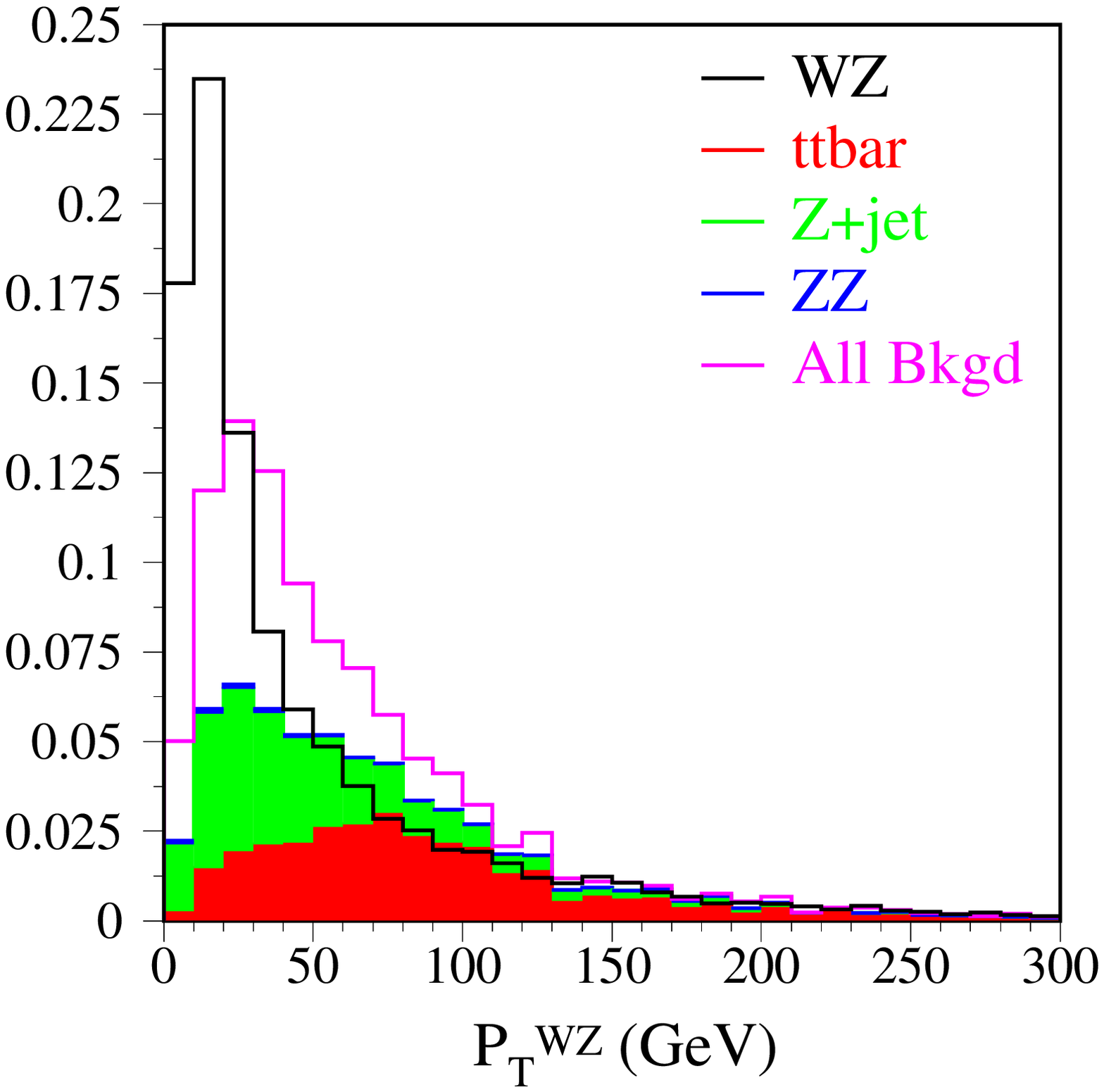,width=8.5cm}
\epsfig{figure=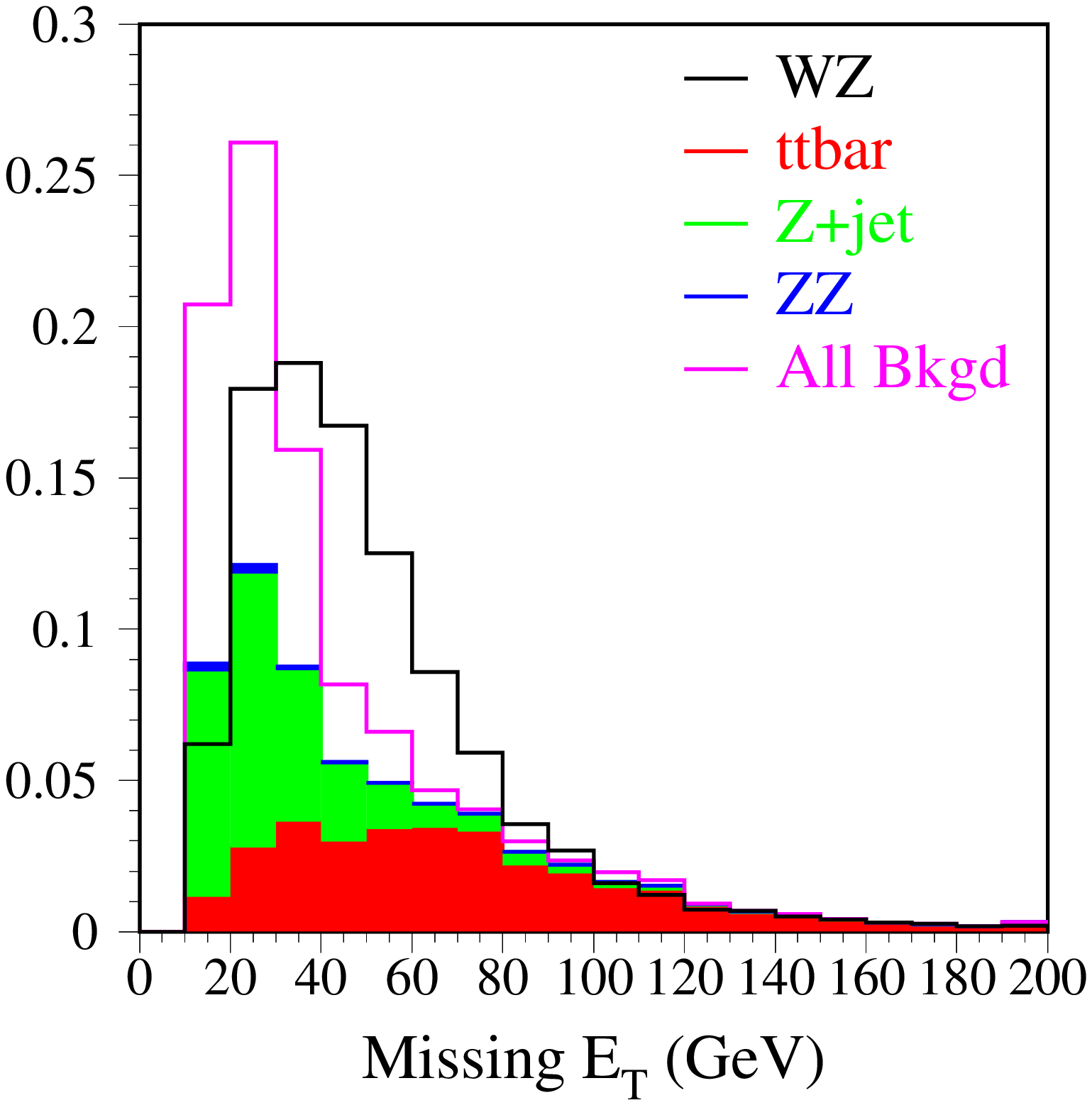,width=8.5cm}
\epsfig{figure=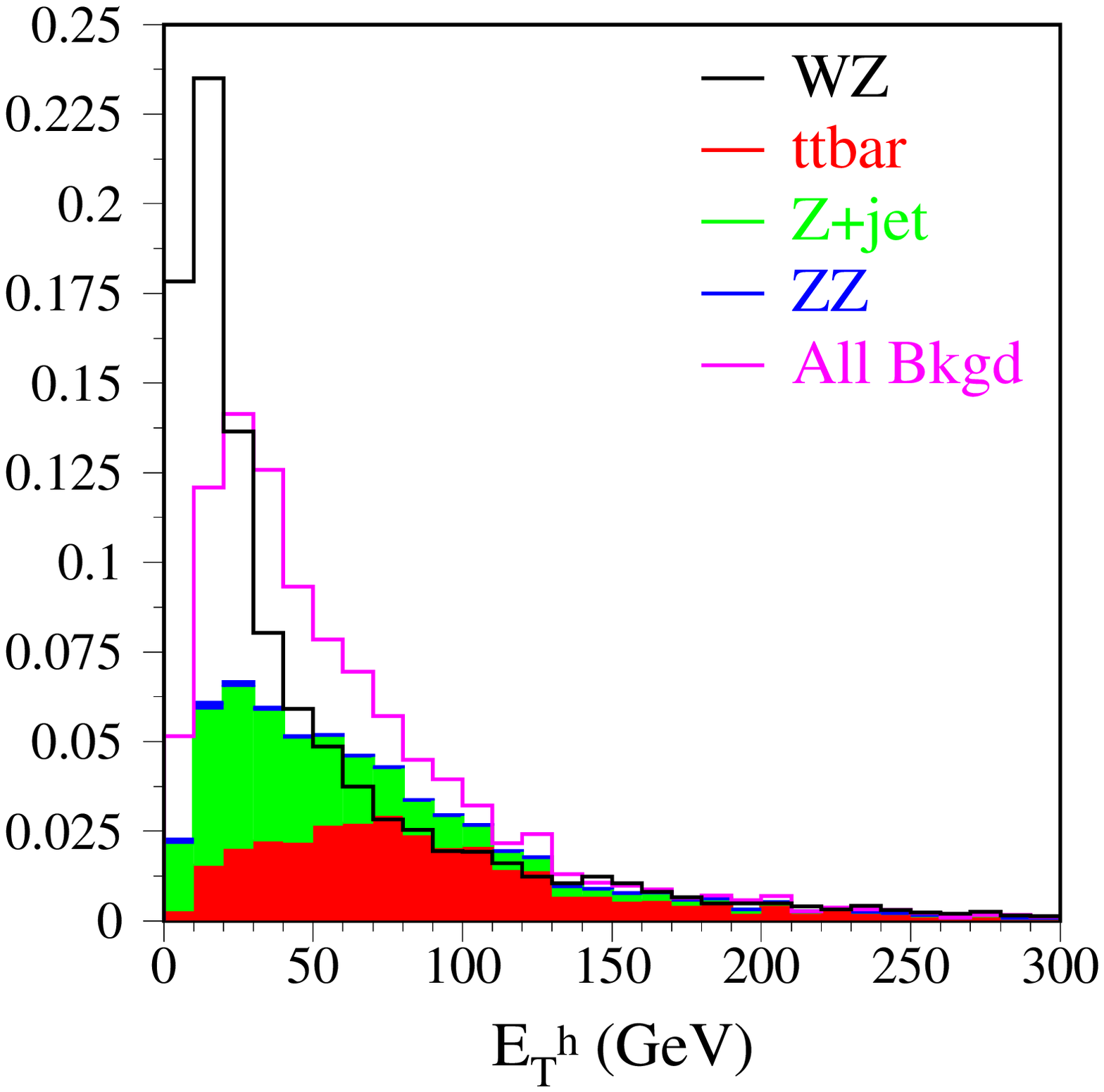,width=8.5cm}
\caption{Distributions of 
the transverse momentum of leptons (top left),
the transverse momentum of the WZ system (top right), 
the missing transverse energy of the event (bottom left) and
the vector sum of transverse momenta from leptons and $MET$ (bottom right).
Among the histograms, black indicates ZW signal events, 
red indicates $t\bar t$, 
green indicates Z plus jets, 
blue indicates $ZZ \rightarrow \ell\ell\ell\ell$ and 
pink indicates a combination of all backgrounds. 
Signal and total reweighted background events are normalized to the same area for comparison.
Major background are stacked indicating the relative contributions.}
\label{fig:var3}
\end{figure}

\begin{figure}
\epsfig{figure=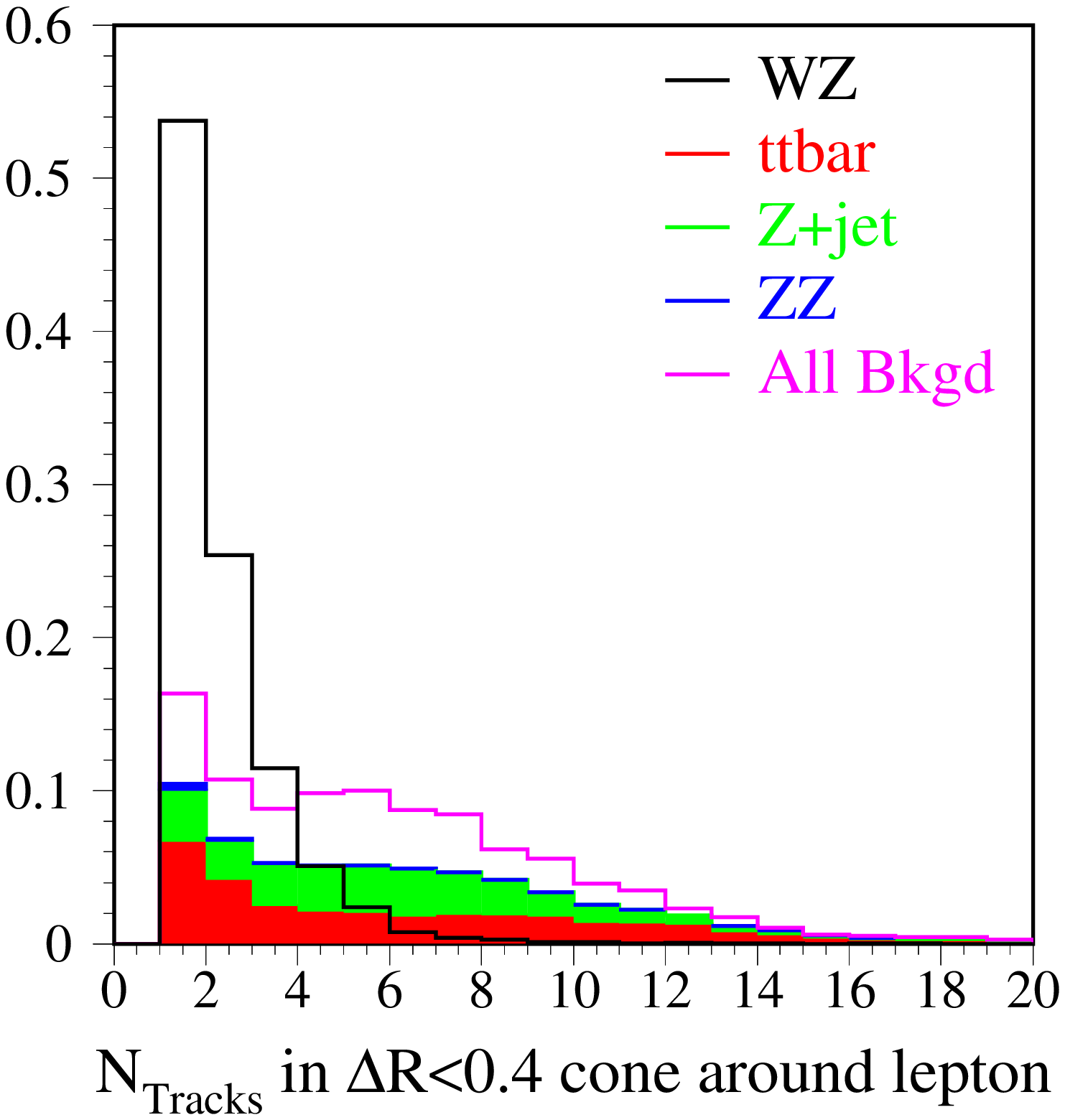,width=8.5cm}
\epsfig{figure=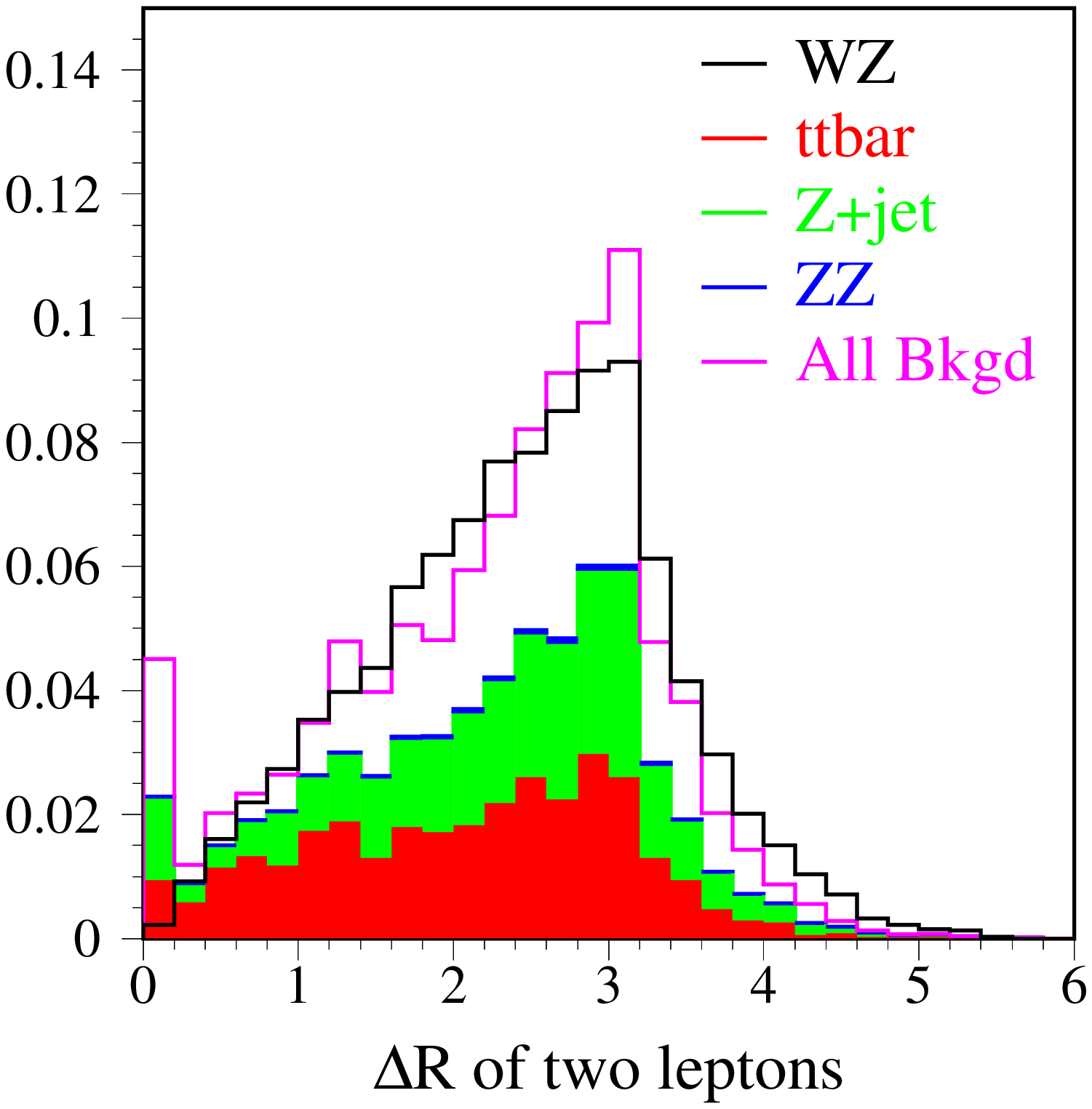,width=8.5cm}
\epsfig{figure=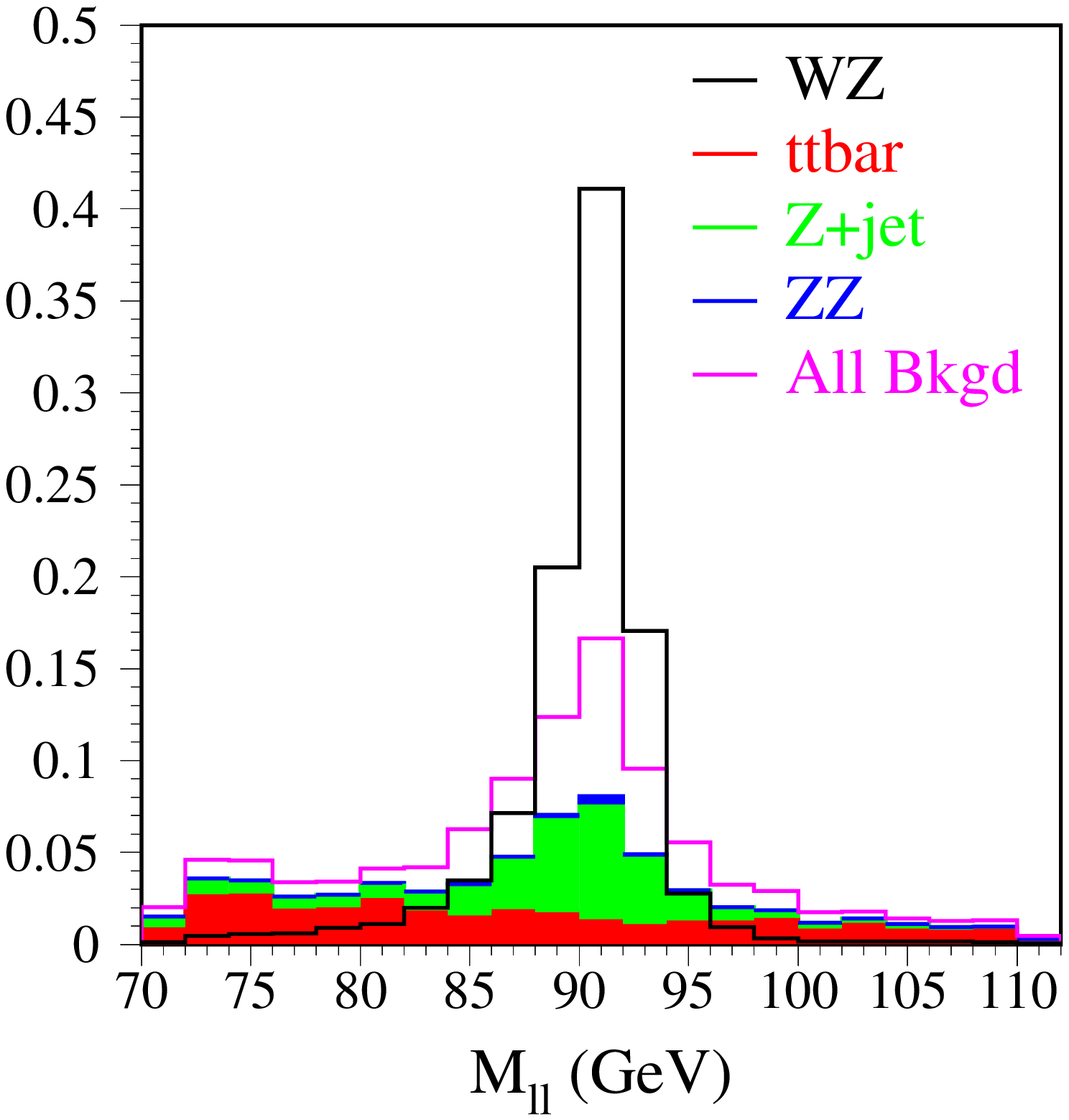,width=8.5cm}
\epsfig{figure=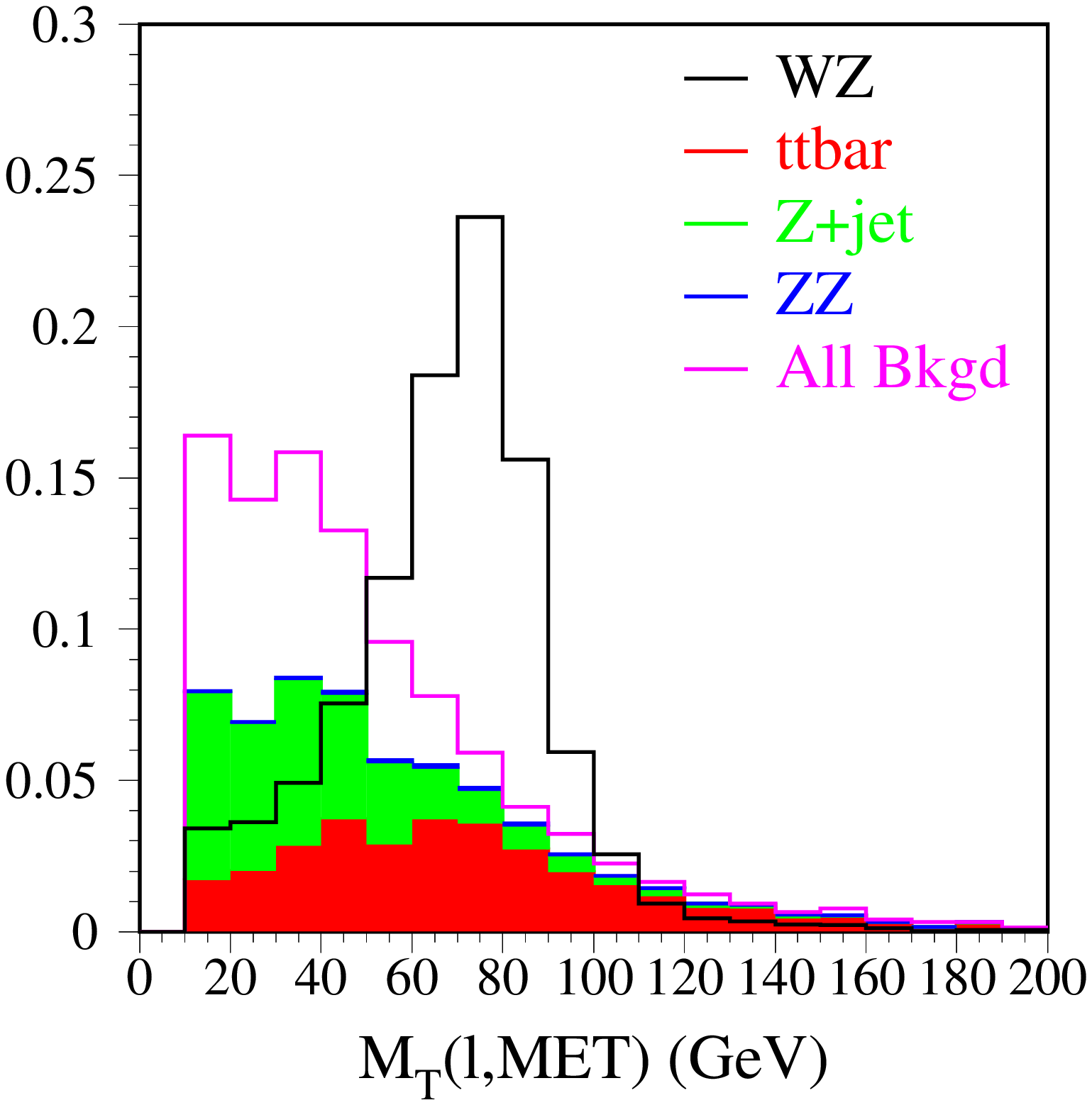,width=8.5cm}
\caption{Distributions of the number of charged tracks around a lepton in a cone of $\Delta R=0.4$ (top left), 
the two lepton separation in $\Delta R$ (top right),
the invariant mass of two leptons (bottom left) and the transverse mass of leptons combined with $MET$ (bottom right).
Among the histograms, black indicates ZW signal events,
red indicates $t\bar t$, 
green indicates Z plus jets, 
blue indicates $ZZ \rightarrow \ell\ell\ell\ell$ and 
pink indicates a combination of all backgrounds. 
Signal and total reweighted background events are normalized to the same area for comparison.
Major background are stacked indicating the relative contributions.}
\label{fig:var4}
\end{figure}

Figure~\ref{fig:var3} and ~\ref{fig:var4} show the same variable distributions with event weighting. 
In these plots, the background histograms are stacked, with areas reflecting the relative weights.
For comparison, total background events and total signal events are
normalized to the same area, respectively.

From the variable distributions (both for signal and background), we found that
a single variable has limited power to separate signal from 
background. But, when combining these variables using ANN or BDT,
the signal and background could be well separated, particularly when the proper
event reweighting algorithm is used in the multivariate analysis.

The BDT program provides a sensitive measure to indicate the signal
and background separation effectiveness of each input variable based on
the Gini index contribution~\cite{bdt1}.
We list the Gini index contributions of the input variables in our analysis 
for both event reweighting and equal weighting cases in Table~\ref{tab:vars}. 
For each variable, a larger Gini index indicates a relatively larger contribution to the overall signal to background separation.
From the Gini index listed in this table we know that
the lepton isolation and mass variables are especially effective at separating the WZ signal
from various background events. However, for equal weighting training, the BDT algorithm
tends to focus on separating the WZ signal from the ZZ background mainly because the
ZZ events dominate the background training sample after the pre-selection.

\section{Event Reweighting Training Technique}

As we mentioned in Section 2, the event weights of various MC processes are
quite different. MC samples with larger event weights represent lower statistics relative to 
cross-section and vice versa.
For instance, the MC $pp \rightarrow ZZ \rightarrow \ell\ell\ell\ell$ sample
has a total of 35700 events (before the pre-selection).  A total cross
section of 18860 fb (NLO) and a four-lepton decay branching ratio of 0.0045 means that
for 1~fb$^{-1}$ integrated luminosity the total number of expected $ZZ \rightarrow \ell\ell\ell\ell$
is about 85.  Thus, the $ZZ$ event weight is 0.0024, as listed in Table~\ref{tab:mc}.
In contrast, the Drell-Yan sample at the Z mass ($pp \rightarrow Z(\ell^+\ell^-)(Mz=81-100~GeV)$)
contains 3.28 million events and yet the NLO cross-section and branching ratio
indicate that 1.85 times as many events (6.05 million)
are expected in $1~fb^{-1}$ integrated luminosity, thus this sample has a weight of 1.85.  

If we treat these MC events from different sources equally using 
conventional training techniques, then multivariate training methods for 
ANNs and BDTs will focus disproportionately on the MC events with lower event weights.
This is because those events have a higher probability of being selected 
for the ANN training relative to their rate of production in the experiment.
In the BDT training process, a sample with larger statistics relative to cross-section
will have relatively larger total event weight. 
This will affect which variables are chosen to be
split in the tree and the value at which the splitting cut occurs.
For example, in our analysis undue emphasis would be put on the variables separating the $ZZ$ background from signal.
To avoid this training prejudice we used event reweighting for the ANN and BDT training. 
As illustrated above, the event weights for different physics processes 
are independent of pre-selection as shown in the Weight definition expression in section 2 and 
Table~\ref{tab:mc}. 
As will be described in the following subsections, 
the sums of all the weights of training signal or background events 
after pre-selection are normalized to 1.

\subsection{BDT Reweighting}

In the BDT training process we start with $N_s$ pre-selected signal and $N_{bg}$ pre-selected background events. 
In the traditional BDT algorithm\cite{bdt1,bdt2} using equal event weighting for training, 
the initial weights of signal and background events are 
$N_s^{-1}$ and $N_{bg}^{-1}$, respectively. The total signal event weight and 
the total background event weight are each normalized to 1.
We implement the event reweighting training technique for BDT by initializing the weights 
of all the training events to the event weights ($WT_s(i),i=1,2,...,N_s$ for signal events;
$WT_{bg}(i),i=1,2,...,N_{bg}$ for background events) listed in Table~\ref{tab:mc}
, then we normalize the total signal
event weight and the total background event weight each to 1.
For signal events, the initial weight for BDT training is
$$
wt_s(i) = WT_s(i)/WTOT_s, i=1,2,...,N_s,
$$
where
$$
WTOT_s = \sum_{i=1}^{N_s} WT_s(i).
$$
For background events, the initial weight for BDT training is
$$
wt_{bg}(j) = WT_{bg}(j)/WTOT_{bg}, j=1,2,...,N_{bg},
$$
where
$$
WTOT_{bg} = \sum_{j=1}^{N_{bg}} WT_{bg}(j).
$$
In our analysis, we have used the $\epsilon-$boost algorithm \cite{bdt1} with 
$\epsilon = 0.01$. For the BDT training we used 1000 tree iterations and 
20 terminal leaves per decision tree.

\subsection{ANN Reweighting}

For conventional ANN training both signal and background events are 
selected randomly with the same probability for each training iteration. 
This is the equal event weighting training technique.
When we have multiple sources of background events with different
production cross-sections from the proton-proton collisions
the equal event weighting training technique may not work well, 
particularly when the number of background events in the
training sample are not proportional to the production cross-sections.
Effectively, the background events with large cross-section are underrepresented, and don't receive appropriate training.
So, we developed the event reweighting algorithm to improve the training
process for ANN. The basic idea of the event reweighting technique is to 
modify the probability of a given event to be selected for the ANN training.
For all the MC events, the reweighting (determination of the probability) 
should be automatically included in the ANN program, which reflects the 
weights of the underlying physics. We briefly describe our algorithm below.

Suppose we have three different background samples, A, B and C. These samples have 
$N_A$, $N_B$ and $N_C$  events, respectively.  
Based on production cross-sections and the
pre-selection efficiencies, we expect the background contributions
from sample A, B, and C are 50\%, 30\% and 20\% respectively.  Thus, 
sample A events should have 50\% probability to be selected for training, 
sample B and C should have 30\% and 20\% probabilities to be selected
for training, respectively.  So, the probabilities of selecting a single training
event in background samples A, B and C are 50\%/$N_A$, 30\%/$N_B$ and 20\%/$N_C$, respectively.

The general algorithm, as implemented in the code is the following,
\begin{itemize}
\item start with weights of training signal and background events listed in 
Table~\ref{tab:mc}, \\
 $wt(j), ~j=1,~2,...,N_i$ (i = signal or background),
\item calculate accumulated weights for event $j$, $wt\_sum(j)$: \\
       $wt\_sum(j) = wt\_sum(j-1) + wt(j),~j=2,...,N$, 
\item generate a random number with uniform distribution in a range of $[0,wt\_sum(N)]$,
\item select an event for the ANN training by minimizing the generated random number $R_n$ and the
accumulated weight $wt\_sum(i)$: min$|R_n-wt\_sum(i)|, ~i=1,...,N$,
\item iterate the above process many times for the ANN training.
\end{itemize}

The event reweighting training technique can be applied to
various ANN algorithms. For this analysis, we used a back-propagation 
neural network with three layers, one input, one hidden and one output layer. 
There are 22 nodes in both the input and hidden layers, 
and there is one output node. The neuron response is a sigmoid function.
The learning rate is $\eta = 0.05$, the momentum is $\alpha=0.07$ and
1,000,000 training cycles are used for the ANN
training. The general description of the ANN
can be found in section 6.8 of the TMVA User guide\cite{tmva}.

\section{Application and Results}

The MC signal $WZ \rightarrow \ell \nu \ell\ell$ and all the background pre-selected events 
are split into two nearly equal samples. Odd and even numbered 
events in each MC process are grouped into sample A and B, respectively. 
Sample A and B are statistically independent.
We use sample A for training and then use sample B for testing.

\begin{figure}
\epsfig{figure=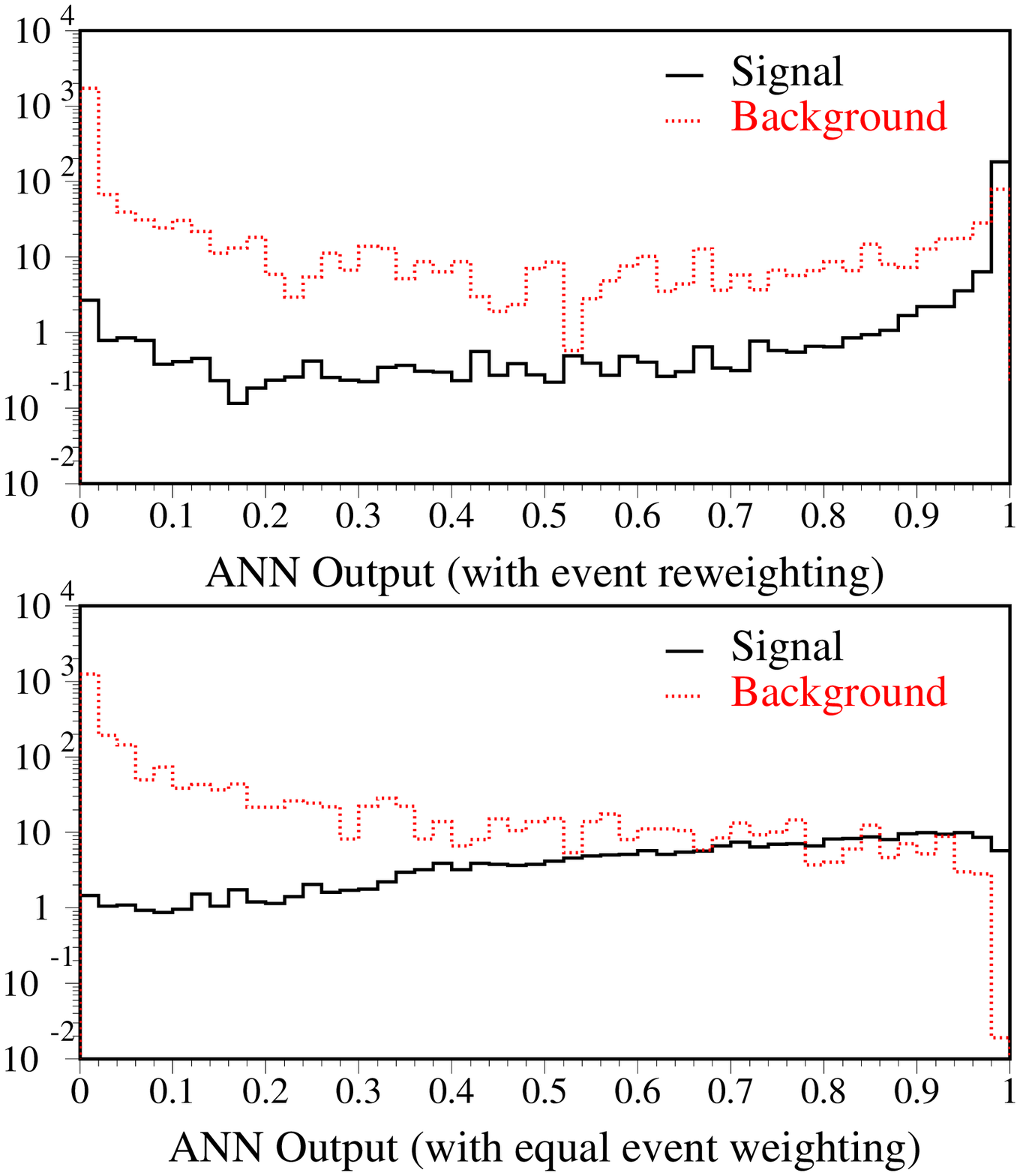,width=14.5cm}
\caption{Distributions of the ANN output for testing samples 
assuming integrated luminosity of 1 fb$^{-1}$.}
\label{fig:output_ann}
\end{figure}

\begin{figure}
\epsfig{figure=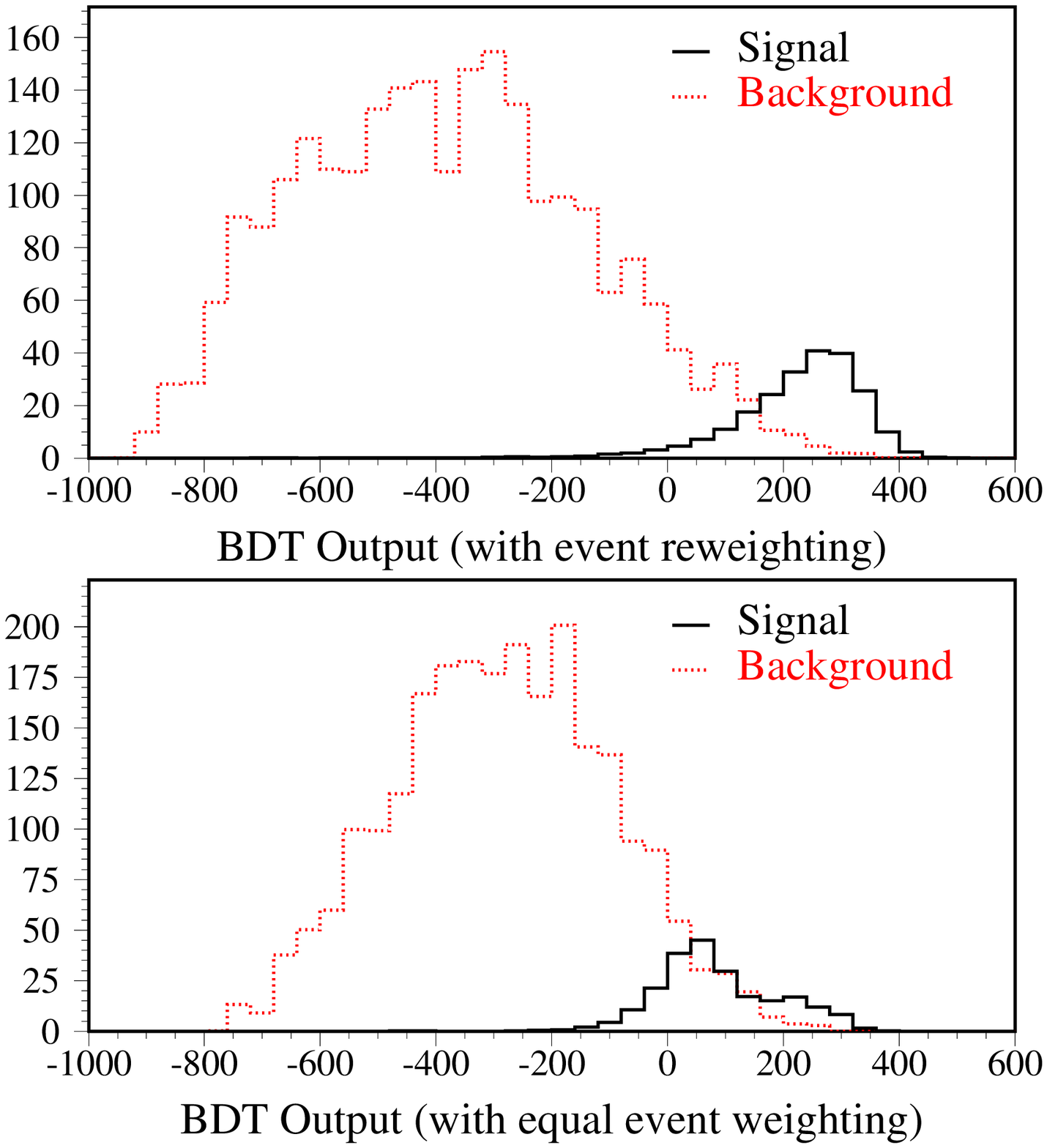,width=14.5cm}
\caption{Distributions of the BDT output for testing samples
assuming integrated luminosity of 1 fb$^{-1}$.}
\label{fig:output_bdt}
\end{figure}

The training sample had 5983 signal and 7907 background 
events.  The testing sample had 5983 signal and 7894 background events.
We performed the ANN and the BDT analysis by using equal weight and reweighting
techniques. We show both results in Figure~\ref{fig:output_ann} and
Figure~\ref{fig:output_bdt}. 
The results shown in these plots are for the testing sample.
Figure~\ref{fig:output_ann} shows the ANN
analysis results and Figure~\ref{fig:output_bdt} shows the BDT results.
In both figures, the top plots show the analysis with the event reweighting
technique and the bottom plots show results with the equal weight technique.
In those plots, the solid histograms are for the signal and the dotted histograms
are for the backgrounds. Both signal and backgrounds are normalized 
to 1 fb$^{-1}$ integrated luminosity.

As we expected, the signal events are mainly distributed in an area close to
1 in the ANN output spectrum, and the background events are distributed
close to 0 (see Figure~\ref{fig:output_ann}). 
The top plot of Figure~\ref{fig:output_ann}, the ANN output spectrum produced
with the reweighting algorithm,
shows that the signal distribution is much sharper around
an ANN output of 1. However, with equal weight training, the signal distribution
near 1 is smeared out. 
As a result, the signal selection efficiency
decreases significantly when using conventional technique of training with equal weights.

Similarly, we observed in Figure~\ref{fig:output_bdt}
that the BDT output with event reweighting training 
has much better signal to background separation power (top plot) compared to
that with equal weighting event training (bottom plot).

By choosing the selection cuts on the ANN or the BDT output spectra, we can
determine the number of selected signal and  background events
as well as the experimental signal to background ratio. Our comparison
of the analysis performance using different training techniques focuses on the
relative difference between signal and background, thus we look at the
number of selected signal events versus background events as the selection cut varies.

\begin{figure}
\epsfig{figure=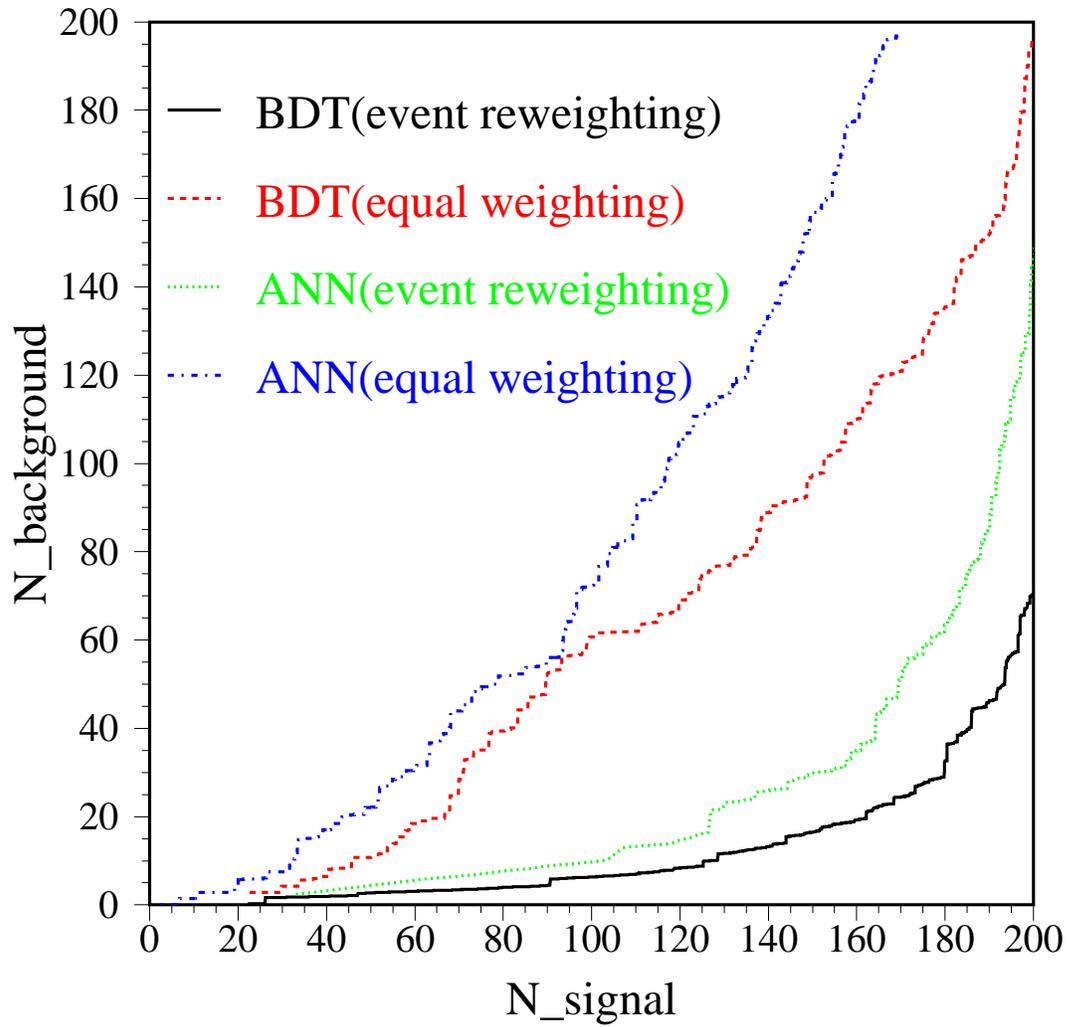,width=14.5cm}
\caption{Number of background events versus number of signal events for testing samples
assuming integrated luminosity of 1 fb$^{-1}$.}
\label{fig:result}
\end{figure}

Comparisons of reweighting and equal weighting techniques
with the ANN and BDT analyses are shown in Figure~\ref{fig:result},
where we plot the expected number of background events versus the number of signal events
corresponding to an integrated luminosity of 1 fb$^{-1}$ by varying the ANN and the BDT
output selection cuts.
The black solid curve represents results from the BDT with event reweighting for training;
the red dashed curve shows results from the BDT with equal event weighting;
the green dotted curve indicates results from the ANN with event reweighting;
and, the blue dash-dotted curve shows results from the ANN with equal event weighting for training.
From these curves we see that, for the same number of signal events selected
using ANN or BDT, using the event reweighting technique gives much lower background event 
contamination compared to equal event weighting.

The numerical comparisons are shown in Table~\ref{tab:comp}. 
We vary the selection cuts on the ANN and BDT output spectra to
keep the same number of signal events and then 
compare the background contamination 
and the ratio of background contamination determined from reweighting and 
equal event weighting techniques. 
Our analysis shows that,
compared with the equal weighting training, 
the ANN and BDT trained with event reweighting reduce the background 
by factors of about $5 \sim 7$ and $6 \sim 10$, respectively. 

\section{Uncertainty Studies}

The reweighting technique will rely on our knowledge of the event production rates
(cross-sections) in the colliders. Thus, it is important to understand the
multivariate training stability with respect to the production cross-section
uncertainties of the MC processes. We looked at BDT training 
to estimate these effects.
We introduced 20\% uncertainties to the event weights for our training samples,
while for the testing sample we have kept the 'correct' event weights.
Compared with the original BDT performance without cross-section uncertainties, 
the relative changes of the BDT performance with 20\%
cross-section uncertainties are less than about 6\%, 
e.g. while keeping the same number of signal events, the background contamination
increased by 4-7\% depending on cuts.
This uncertainty is well within the 15-25\% relative Root-Mean-Squared (RMS) 
errors of background efficiencies which will be described below. 

\begin{figure}
\epsfig{figure=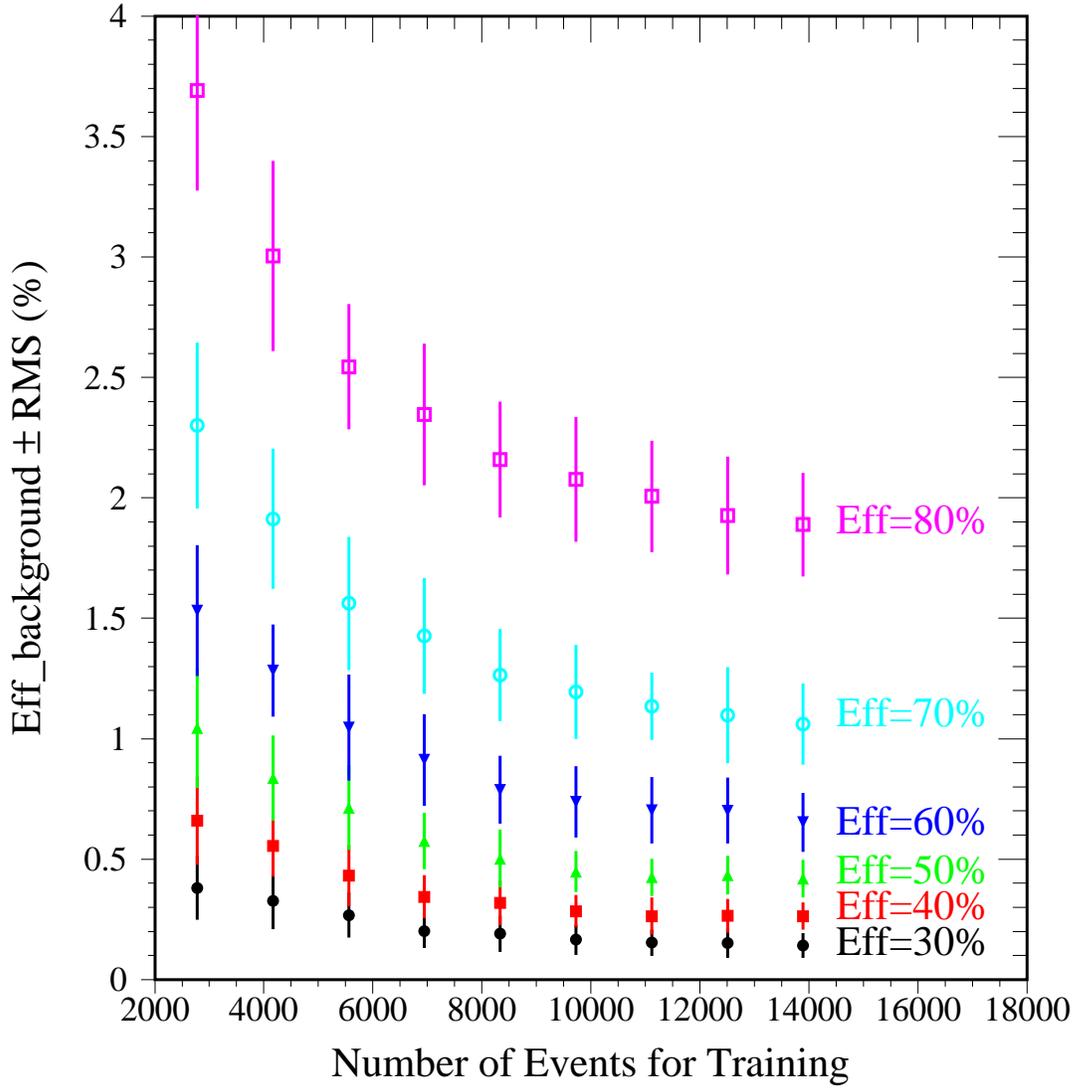,width=14.5cm}
\caption{Background efficiencies versus number of training events for various
signal efficiencies. 
Training events are selected randomly but the testing sample is fixed 
for the comparison. The BDT training-testing process is repeated 50 times
to obtain average background efficiencies and RMS for a set of fixed signal
efficiencies.}
\label{fig:effbg}
\end{figure}

Generally, a large training sample is desired for training in multivariate algorithms.
It is important to understand what is a sufficient number of training events such that
the background efficiency is insensitive to training sample size.
To this end, we studied the background efficiencies and RMS errors
as a function of the number of training events for a set of fixed signal efficiencies 
as shown in Figure~\ref{fig:effbg}.
The training events are selected randomly with replacement from the training sample
with 5983 signal and 7907 background events for the BDT training, and
a statistically independent MC sample with 5983 signal and 7894 background events
is used for testing. For each training point,
the BDT training-testing process is repeated 50 times with a different set of
training samples and a fixed testing sample to obtain the average background efficiencies and
RMS errors for a given set of signal efficiencies.
Figure~\ref{fig:effbg} indicates that fewer MC events for training
will result in larger background contaminations, presumably
because the number of MC events is insufficient to fully train the BDTs.
We also note that with at least 10000 training events the background efficiency
becomes relatively stable.  We have used about 14000 events for training, thus we expect
the bias due to training sample size should be small.

\section{Conclusions}

We have developed and tested an event reweighting technique to be used when training
multivariate pattern recognition processes.
This technique is necessary to train the pattern recognition in an unbiased way,
particularly for multi-background processes with limited MC statistics.
For the  ATLAS $WZ \rightarrow \ell \nu \ell\ell$ analysis, with large background
contributions from different physics processes, 
we found that for good performance using the ANN and the BDT analysis
one should employ event reweighting in the training process. 

\section{Acknowledgments}

We wish to express our gratitude to the ATLAS Collaboration for their excellent
work on the Monte Carlo simulation and physics analysis software.
We also would like to thank Byron P. Roe and Ji Zhu for useful discussions.
This work is supported by the Department of Energy (DE-FG02-95ER40899) 
of the United States.









\begin{table}
\begin{center}
{\scriptsize {
\begin{tabular}{|l|c|c|c|r|r|r|r|} \hline
MC Process & $\sigma_{MC}$(fb) & K & Br & $N_{MC}$ & $N_{precut}$ & $N_{exp} / fb^{-1}$ & Weight \\ \hline\hline
$pp \rightarrow W^+Z \rightarrow \ell^+\nu\ell^+\ell^-$   &     0.3673E+05 &    1.0 &    0.0144 &      26550 &    6848 &    136.4 &     0.0199 \\ \hline
$pp \rightarrow W^-Z \rightarrow \ell^-\bar{\nu}\ell^+\ell^-$   &     0.2099E+05 &    1.0 &    0.0144 &      17450 &    5118 &     88.7 &     0.0173 \\ \hline
$pp \rightarrow Z/\gamma \rightarrow \ell^+\ell^-$  &    0.8910E+06 &    1.5 &    0.0672 &     999742 &     111 &     10.0 &     0.0898 \\ \hline
$pp \rightarrow Z(e^+e^-)+jet(Ejet=10-20 ~GeV) $   &        0.1360E+08 &    1.3 &    0.0336 &     597281 &       0 &      0.0 &     0.9946 \\ \hline
$pp \rightarrow Z(e^+e^-)+jet(Ejet=20-40 ~GeV) $   &        0.8670E+07 &    1.3 &    0.0336 &     398697 &       0 &      0.0 &     0.9499 \\ \hline
$pp \rightarrow Z(e^+e^-)+jet(Ejet=40-80 ~GeV) $   &        0.4120E+07 &    1.3 &    0.0336 &     397524 &       0 &      0.0 &     0.4527 \\ \hline
$pp \rightarrow Z(e^+e^-)+jet(Ejet=80-120 ~GeV)$   &        0.8270E+06 &    1.3 &    0.0336 &     397009 &       0 &      0.0 &     0.0910 \\ \hline
$pp \rightarrow Z(e^+e^-)+jet(Ejet>120 ~GeV)   $   &        0.3830E+06 &    1.3 &    0.0336 &     198652 &       0 &      0.0 &     0.0842 \\ \hline
$pp \rightarrow Z(\tau^+\tau^-)+jet(Ejet=10-20 ~GeV) $   &  0.1360E+08 &    1.3 &    0.0336 &     598783 &       0 &      0.0 &     0.9921 \\ \hline
$pp \rightarrow Z(\tau^+\tau^-)+jet(Ejet=20-40 ~GeV) $   &  0.8670E+07 &    1.3 &    0.0336 &     399076 &       0 &      0.0 &     0.9490 \\ \hline
$pp \rightarrow Z(\tau^+\tau^-)+jet(Ejet=40-80 ~GeV) $   &  0.4120E+07 &    1.3 &    0.0336 &     398972 &       0 &      0.0 &     0.4511 \\ \hline
$pp \rightarrow Z(\tau^+\tau^-)+jet(Ejet=80-120 ~GeV)$   &  0.8270E+06 &    1.3 &    0.0336 &     396671 &       0 &      0.0 &     0.0911 \\ \hline
$pp \rightarrow Z(\tau^+\tau^-)+jet(Ejet>120 ~GeV)   $   &  0.3830E+06 &    1.3 &    0.0336 &     199046 &       0 &      0.0 &     0.0840 \\ \hline
$pp \rightarrow Z(\mu^+\mu^-)+jet(Ejet=10-20 ~GeV)    $   &  0.1360E+08 &    1.3 &    0.0336 &    2996413 &     492 &     97.5 &     0.1983 \\ \hline
$pp \rightarrow Z(\mu^+\mu^-)+jet(Ejet=20-40 ~GeV)    $   &  0.8670E+07 &    1.3 &    0.0336 &    1995792 &     789 &    149.7 &     0.1898 \\ \hline
$pp \rightarrow Z(\mu^+\mu^-)+jet(Ejet=40-80 ~GeV)    $   &  0.4120E+07 &    1.3 &    0.0336 &    1189793 &    1516 &    229.3 &     0.1513 \\ \hline
$pp \rightarrow Z(\mu^+\mu^-)+jet(Ejet=80-120 ~GeV)   $   &  0.8270E+06 &    1.3 &    0.0336 &     397856 &    1105 &    100.3 &     0.0908 \\ \hline
$pp \rightarrow Z(\mu^+\mu^-)+jet(Ejet>120 ~GeV)      $   &  0.3830E+06 &    1.3 &    0.0336 &     199832 &    1133 &     94.9 &     0.0837 \\ \hline
$pp \rightarrow Z(\ell^+\ell^-)(Mz=30-81 ~GeV)     $   &  0.4220E+07 &    1.3 &    0.1010 &    1000000 &      16 &      8.9 &     0.5541 \\ \hline
$pp \rightarrow Z(\ell^+\ell^-)(Mz=81-100 ~GeV)    $   &  0.4610E+08 &    1.3 &    0.1010 &    3284999 &     406 &    748.1 &     1.8426 \\ \hline
$pp \rightarrow Z(\ell^+\ell^-)(Mz>100 ~GeV)       $   &  0.1750E+07 &    1.3 &    0.1010 &     971000 &     271 &     64.1 &     0.2366 \\ \hline
$pp \rightarrow Z\mu\mu ~(M_{inv}>150~GeV)$            &  0.1750E+07 &    0.8 &    0.0336 &      43000 &      33 &     36.1 &     1.0940 \\ \hline
$pp \rightarrow Z\mu\mu Jet $       &        0.8270E+06 &    0.8 &    0.0336 &      35000 &      20 &     12.7 &     0.6351 \\ \hline
$pp \rightarrow Zee~(Pt>100 ~GeV)$       &       0.8270E+06 &    0.8 &    0.0336 &      46000 &      11 &      5.3 &     0.4833 \\ \hline
$pp \rightarrow Z\mu\mu~(Pt>100 ~GeV)$     &       0.8270E+06 &    0.8 &    0.0336 &      33000 &      42 &     28.3 &     0.6736 \\ \hline
$pp \rightarrow Z\tau\tau~(Pt>100 ~GeV)$   &       0.8270E+06 &    0.8 &    0.0003 &      32000 &      41 &      0.3 &     0.0069 \\ \hline
$pp \rightarrow t\bar{t}  $    &      0.7590E+06 &    1.0 &    0.5550 &     604750 &    1071 &    746.0 &     0.6966 \\ \hline
$pp \rightarrow Z\gamma~(Pt>25 ~GeV)     $ &         0.4510E+05 &    1.0 &    0.0672 &      46800 &      43 &      2.8 &     0.0648 \\ \hline
$pp \rightarrow W^+W^- \rightarrow e^+\nu e^-\bar{\nu}       $  &       0.1133E+06 &    1.0 &    0.0120 &      41950 &       9 &      0.3 &     0.0324 \\ \hline
$pp \rightarrow W^+W^- \rightarrow e^+\nu \mu^-\bar{\nu}     $  &       0.1133E+06 &    1.0 &    0.0120 &      45900 &      22 &      0.7 &     0.0296 \\ \hline
$pp \rightarrow W^+W^- \rightarrow e^+\nu \tau^-\bar{\nu}    $  &       0.1133E+06 &    1.0 &    0.0120 &      71000 &       7 &      0.1 &     0.0191 \\ \hline
$pp \rightarrow W^+W^- \rightarrow \mu^+\nu e^-\bar{\nu}     $  &       0.1133E+06 &    1.0 &    0.0120 &      47000 &      18 &      0.5 &     0.0289 \\ \hline
$pp \rightarrow W^+W^- \rightarrow \mu^+\nu \mu^-\bar{\nu}   $  &       0.1133E+06 &    1.0 &    0.0120 &      48950 &      30 &      0.8 &     0.0278 \\ \hline
$pp \rightarrow W^+W^- \rightarrow \mu^+\nu \tau^-\bar{\nu}  $  &       0.1133E+06 &    1.0 &    0.0120 &      44000 &       8 &      0.2 &     0.0309 \\ \hline
$pp \rightarrow W^+W^- \rightarrow \tau^+\nu e^-\bar{\nu}    $  &       0.1133E+06 &    1.0 &    0.0120 &      47700 &       2 &      0.1 &     0.0285 \\ \hline
$pp \rightarrow W^+W^- \rightarrow \tau^+\nu \mu^-\bar{\nu}  $  &       0.1133E+06 &    1.0 &    0.0120 &      45800 &       8 &      0.2 &     0.0297 \\ \hline
$pp \rightarrow W^+W^- \rightarrow \tau^+\nu\tau^-\bar{\nu}  $  &       0.1133E+06 &    1.0 &    0.0120 &      34850 &       0 &      0.0 &     0.0390 \\ \hline
$pp \rightarrow ZZ \rightarrow \ell^+\ell^-\ell^+\ell^-      $  &       0.1886E+05 &    1.0 &    0.0045 &      35700 &    8597 &     20.4 &     0.0024 \\ \hline
\end{tabular}
}}
\end{center}
\caption{Breakdown of MC samples used for ZW analysis.}
\label{tab:mc}
\end{table}

\begin{table}
\begin{center}
{\small {
\begin{tabular}{|l|r|r|} \hline
Training Variables & \multicolumn{2}{|c|}{Gini Index Contribution(\%)} \\ \cline{2-3}
                   & Event Reweighting & Equal Weighting \\ \hline
$P_T(Z\rightarrow \ell^-)$                      &  2.41 &  1.46 \\ \hline
$N_{trk}^{iso}$ (tracks around $Z\rightarrow \ell^-$ in $\Delta R < 0.4$ cone)         &  4.53 &  2.13 \\ \hline
$P_T(Z\rightarrow \ell^+)$                      &  1.93 &  1.34 \\ \hline
$N_{trk}^{iso}$ (tracks around $Z\rightarrow \ell^+$ in $\Delta R < 0.4$ cone)         &  7.65 &  2.49 \\ \hline
$P_T(W^\pm\rightarrow \ell^\pm)$                &  4.16 &  3.01 \\ \hline

$\sum P_T^{iso}$ (tracks around $W^\pm\rightarrow \ell^\pm$ in $\Delta R < 0.4$ cone)   & 11.88 & 11.80 \\ 
\hline 
$N_{trk}^{iso}$ (tracks around $W^\pm\rightarrow \ell^\pm$ in $\Delta R < 0.4$ cone)      & 20.56 & 14.56 \\ \hline
$\sum E_T^{iso}$ (jets around $W^\pm\rightarrow \ell^\pm$ in $\Delta R < 0.4$ cone) &  2.07 &  5.83 \\ \hline
$f_\ell^{iso}$                                            &  2.05 &  4.57 \\ \hline
$\Delta A(Z\rightarrow \ell^+,W^\pm\rightarrow \ell^\pm)$ & 3.26 & 2.73 \\ \hline

$\Delta R(Z\rightarrow \ell^+,W^\pm\rightarrow \ell^\pm)$ & 2.63 & 2.49 \\ \hline
$\Delta A(Z\rightarrow \ell^-,W^\pm\rightarrow \ell^\pm)$ & 4.17 & 3.12 \\ \hline
$\Delta R(Z\rightarrow \ell^-,W^\pm\rightarrow \ell^\pm)$ & 3.05 & 3.07 \\ \hline
$MET$-missing transverse energy          & 3.90 & 10.26 \\ \hline
$P_T(WZ)$                                & 1.59 & 3.88 \\ \hline

$M_{\ell\ell}$                           & 9.70 & 4.22 \\ \hline
$M_T(\ell,MET)$                          & 7.55 & 6.67 \\ \hline
$H_t$                                    & 0.91 & 0.94 \\ \hline
$\sum E_T^{jet}$                         & 0.91 & 1.73 \\ \hline
$E_T^h$                                  & 0.72 & 6.03 \\ \hline

$MET/\sqrt{\sum_i E_T(i)}$               & 0.95 & 4.10 \\ \hline
$E_T^{recoil}$                           & 3.42 & 3.58 \\ \hline

\end{tabular}
}}
\end{center}
\caption{Gini index contributions of input variables for BDT training using
event reweighting and equal weighting techniques. $Z\rightarrow \ell$ means lepton
decays from Z and $W\rightarrow \ell$ means lepton decays from W.}
\label{tab:vars}
\end{table}

\begin{table}
\begin{center}
\begin{tabular}{|l|r|r|r|r|r|r|} \hline
$N_{signal}$                  &  60   &  80   &  100 &   120 &   140  &   160  \\ \hline\hline

$N_{bg1}$ for ANN-equal-weighting &  30.5 &  51.9 & 72.4 & 104.7 & 133.3  & 177.6  \\ \hline
$N_{bg2}$ for ANN-event-reweighting &   5.8 &   7.7 &  9.8 &  14.7 &  25.9  &  34.9  \\ \hline
$Ratio=N_{bg1}/N_{bg2}$ for ANN&   5.3 &   6.7 &  7.4 &   7.1 &   5.1  &   5.1  \\ \hline\hline

$N_{bg3}$ for BDT-equal-weighting &  18.5 &  39.4 & 60.7 &  69.1 &  88.9  & 110.1  \\ \hline 
$N_{bg4}$ for BDT-event-reweighting &   3.1 &   4.0 &  6.3 &   8.4 &  13.2  &  19.3  \\ \hline
$Ratio=N_{bg3}/N_{bg4}$ for BDT&   6.0 &   9.9 &  9.6 &   8.2 &   6.7  &   5.7  \\ \hline\hline

\end{tabular}
\end{center}
\caption{Number of background events ($N_{bg}$) versus number of signal events ($N_{signal}$) 
using the ANN and BDT discriminating algorithms with equal weighting and event reweighting training techniques.}
\label{tab:comp}
\end{table}



\end{document}